\documentclass[prb, preprint, unsortedaddress, showpacs, nofootinbib, eqsecnum]{revtex4}

\usepackage{amsfonts, amsbsy, amssymb, amsmath, graphicx, color,  float}

\newtheorem{definition}{Definition}
\newtheorem{assumption}{Assumption}

\restylefloat{figure}
\newcommand{\rmd}{{\rm d}}

\newcommand{\calM}{\mathcal{M}}

\newcommand{\calP}{\mathcal{P}}

\newcommand{\sbar}{\overline{s}}

\newcommand{\Int}[3]{\int_{#2}^{#3}\rmd {#1} \;}

\newcommand{\bz}{\boldsymbol{z}}

\newcommand{\bbP}{\mathbb{P}}

\begin{document}


\title{Bifurcations of Normally Hyperbolic Invariant Manifolds and Consequences for  
Reaction Dynamics
}



\author{Fr\'ed\'eric A. L. Maugui\`{e}re}
\email[]{frederic.mauguiere@bristol.ac.uk}
\affiliation{School of Mathematics  \\
University of Bristol\\Bristol BS8 1TW\\United Kingdom}

\author{Peter Collins}
\affiliation{School of Mathematics  \\
University of Bristol\\Bristol BS8 1TW\\United Kingdom}

\author{Gregory S. Ezra}
\email[]{gse1@cornell.edu}
\affiliation{Department of Chemistry and Chemical Biology\\
Baker Laboratory\\
Cornell University\\
Ithaca, NY 14853\\USA}

\author{Stephen Wiggins}
\email[]{stephen.wiggins@mac.com}
\affiliation{School of Mathematics \\
University of Bristol\\Bristol BS8 1TW\\United Kingdom}


\date{\today}

\begin{abstract}
In this paper we study the breakdown of normal hyperbolicity and its 
consequences for reaction dynamics; in particular, 
the dividing surface, the flux through the dividing surface (DS), 
and the gap time distribution. Our approach is to study these questions using simple, 
two degree-of-freedom Hamiltonian models where calculations for 
the different geometrical and dynamical quantities can be carried out exactly. 
For our examples, we show that resonances within the 
normally hyperbolic invariant manifold may, or may not, lead to a `loss of normal hyperbolicity'.
Moreover, we show that the onset of such resonances 
results in a change in topology of the dividing surface, 
but does not affect our ability to define a DS.
The flux through the DS 
varies continuously with energy, even as the energy is varied in such a way that normal hyperbolicity is lost. 
For our examples the gap time distributions exhibit singularities 
at energies corresponding to the existence of homoclinic orbits in the DS, but 
these singularities are not associated with loss of normal hyperbolicity. 
\end{abstract}


\maketitle


\section{Introduction}
\label{sec:intro}

Transition state theory  has played, and continues to play, a fundamental role in how we think of chemical reactions. 
There are many excellent reviews of the subject; see, for example, refs 
\onlinecite{Pechukas81,PollakTalkner,LaidlerKing,Garrett,Petersson}. 
 The work of Wigner \cite{Wigner1938} highlighted the notion of the transition state as 
 a dividing surface (DS) in phase space, having  the property that trajectories 
 crossed this surface (only once) in their evolution from reactants to products.
 This phase space point of view was relatively undeveloped for many years until the 1970's 
 when such a dynamical theory was fully realised for two degrees of freedom (DoF)
in  the work of Pechukas, Pollak and Child \cite{child1980analytical,pollak1980classical,
polChilPec1980,pechukas1979classical,pollak1978transition,pechukas1977trapped,pollak1979unified}.  
In their work they realised the construction of such a DS having the `no recrossing' property, 
and a central role was played by a saddle-like (in terms of stability) 
periodic orbit (PO) which serves as the `anchor' for the definition of the DS. 
This PO was called a
periodic orbit dividing surface (PODS). 
(This choice of name was perhaps unfortunate as the PO itself is not the DS [in phase space] but rather 
the \emph{boundary} of  the DS.)
The work of Pechukas, Pollak and Child just cited was
the first example of the use of a normally hyperbolic invariant manifold (NHIM) 
in the context of transition state theory (although see also ref.\ \onlinecite{Devogelaere55}).

For more than two DoF a suitable generalisation of a saddle like (or `hyperbolic') 
periodic orbit was required, and this was realised in the notion 
of a normally hyperbolic invariant manifold (NHIM). Explicit discussion of NHIMs 
in the context of  reaction theory was in refs \onlinecite{wiggins90, Gillilan91}
(see also refs \onlinecite{Gillilan89,gillilan1990invariant}). 
However, the usefulness of NHIMs for the definition of phase space dividing surfaces 
was fully realised only after the development of methods  for computing NHIMS and their associated DS.
In particular, NHIMs could be computed using the classical Poincar\'e-Birkhoff normal form procedure, 
and its quantum analog \cite{wiggins2001impenetrable,
uzer2002geometry,waalkens2007wigner}. However, the Poincar\'e-Birkhoff normal  
form procedure is local in nature since it involves a Taylor expansion about an 
appropriate saddle point.  Practically, the expansion must be truncated at 
an appropriate order so that its ability to describe geometrical structures and dynamics  
at a desired accuracy must be assessed (and this is, typically, problem dependent). 
Very broadly speaking, the NHIM, and its associated local dynamics, are 
accurately described by the truncated normal form for energies `close'
to the energy of the saddle point. To date, our knowledge of NHIMs for specific problems 
relies heavily on the techniques used to compute them, in particular normal form theory. 
The local nature of these techniques makes NHIMs tools of limited 
utility in terms of our ability describe reaction dynamics
at energies well above the reaction threshold. 
In particular, very little is known about how the NHIM behaves as we increase the energy 
above that of the saddle. We expect the normal form to `break down' (in some not very precisely defined sense). 
However, the breakdown of the normal form and the breakdown of the NHIM are  
two separate issues. Questions concerning the existence and properties of the NHIM 
can in principle be considered independently of the methods used to compute the NHIM,
when normal form approximations may  no longer apply.

In this paper we study the behaviour of NHIMs as the energy above the 
saddle point is increased, and  the dependence on energy of associated 
quantities related to reaction dynamics.
Our strategy will be to avoid the use of normal forms by considering 
simple models that allow us to explicitly compute the NHIM, the DS, and 
the  geometrical properties
associated with reaction dynamics. 
We can view our approach as a decoupling of normal form issues 
from the dynamics of reaction. 
We then  ask the question: ``what happens to a NHIM as parameters 
(the energy being a very important parameter) are varied''?  
In other words, we  study `bifurcation of NHIMs'. 
But, more specifically for reaction dynamics, 
we want to understand how such  bifurcation of NHIMs affects the 
various quantities (described below) used to describe reaction dynamics.

First, we need to describe some background regarding what we mean by the `bifurcation of a NHIM'
and then discuss our approach for studying these phenomena.  
Roughly speaking (for details, see Section \ref{sec:NHIMmodels})  a normally hyperbolic invariant manifold has the property that the {\em linearized} growth rates  normal to the manifold dominate the {\em linearized} growth rates tangent to the manifold.  The nature of  the dynamics within the invariant manifold can be arbitrary, provided the growth rate conditions are satisfied. Hence, there are several possibilities to consider:
\begin{itemize}

\item Bifurcation within the NHIM, but with the normal hyperbolicity conditions not affected.

\item Breakdown of the normal hyperbolicity conditions.

\item A combination of the above two phenomena -- 
bifurcation within the NHIM and breakdown of the normal hyperbolicity conditions.

\end{itemize}

How bifurcation within the NHIM occurs and how the normal hyperbolicity conditions can break down
will be discussed in detail when we consider explicit models in Section \ref{sec:models}.   
Briefly, there are two ways in which such bifurcations can occur. 
One is when a parameter in the Hamiltonian is varied, 
and the other is when the energy itself is varied. 
(We will be considering only Hamiltonian systems.)
Of course, energy is a parameter, but its importance in understanding the dynamics 
is such that we  highlight its effects explicitly.

Typically in bifurcation theory simple `normal forms' are developed that embody the 
essential features of a particular bifurcation. We will follow a similar approach here.  
Our models will be two degree-of-freedom, separable, Hamiltonian systems 
where we have explicit control and understanding of the dynamics in such 
a way that we can model the three possibilities noted above.  
Modelling the three possibilities for separable, 
two DoF Hamiltonian systems is of course fairly straightforward, 
but that is the point. Our goal is to determine the effect that 
the three scenarios outlined above have on quantities that are 
used to describe reaction dynamics.  
In particular, we want to consider the effect on:
\begin{itemize}

\item The  dividing surface (DS) separating reactants and products as a function of energy 
(the DS is defined and explained in Section \ref{subsec:DS}).

\item The flux through the DS as a function of energy.

\item The gap time distribution as a function of energy 
(these quantities are defined and explained in Section \ref{sec:volumes}).

\end{itemize}

Despite the fact that our examples are very simple (completely integrable 2 DoF Hamiltonian systems),
they allow for exact calculations and therefore provide benchmarks
against which more complex examples can be compared.  
In particular, we note 
a previous study by Li and coworkers \cite{li2006definability} which suggests that the breakdown 
of normal hyperbolicity of the NHIM is responsible for the non definability of a 
relevant DS at high energy above the reaction threshold.
This  conclusion is contrary to that reached in the study of our model examples (see below).
A study of Allahem and Bartsch \cite{Allahem12} asserts 
that the transition state loses, and then regains normal hyperbolicity as energy is varied. 
A  study of  Inarrea et al \cite{Inarrea11} concludes that even though bifurcation 
of the NHIM can occur, dividing surfaces  can still be defined. 
These three studies are concerned with  Hamilltonian models that are 
much more complex
than ours (although still 2 DoF).
We also note an interesting study by Yang \cite{yang09}, 
although the models studied there are {\em not} Hamiltonian.

The organisation of this paper is as follow. We review the concepts of NHIMs in Section
\ref{sec:NHIMmodels} where we discuss the definition of normal hyperbolicity in detail. 
In Section \ref{sec:gen_models} we discuss the general form for the two DoF models that we study. 
These models allow us to explicitly compute NHIMs and quantify their geometrical and stability properties. 
In Section \ref{sec:volumes} we discuss  concepts of phase space volumes, gap times, and reaction rates 
that are relevant for the present study. 
In Section \ref{sec:models} we describe the two models for which we will 
carry out explicit calculations.  In Section \ref{subsec:DS} 
we discuss dividing surfaces and compute the flux through the 
dividing surfaces for our two examples, while Section \ref{subsec:gapdistrib} 
deals with the computation of the gap time distribution.  
In Section \ref{subsecnormalhyperbol} we discuss the loss 
of normal hyperbolicity and its consequences for the DS, flux, and gap 
time distribution in our examples. In Section \ref{sec:concl} we present our conclusions.

\newpage

\section{Normally Hyperbolic Invariant Manifolds and Models for Their Bifurcation}
\label{sec:NHIMmodels}

The notion of a {\em normally hyperbolic invariant manifold} (or NHIM) 
is by now  a standard concept and tool in dynamical systems theory that is playing an increasingly 
important role in a variety of applications. The theoretical framework developed over the course of 
many  years, beginning in the early part of the 20th century
and reached a mature form in the works of Fenichel \cite{Fenichel1971,Fenichel1974,Fenichel1977} 
and Hirsch, Pugh, and Shub \cite{hirsch1970invariant}. 
In ref.\ \onlinecite{wiggins1994normally} a (relatively) elementary exposition of Fenichel's approach to NHIMs 
is given with some discussion of applications.

We begin by defining  the notion of a normally hyperbolic invariant manifold (NHIM) 
in a continuous time setting  (i.e. the setting of autonomous ordinary differential equations, rather than maps).  
There are numerous definitions that appear throughout the literature. 
We take the  definition from ref.\ \onlinecite{Delshams2012}.  
From an `applied' point of view the definition may appear somewhat abstract,
leading to difficulties understanding how the definition can be applied in concrete settings.
However, this level of abstraction actually 
provides a great deal of flexibility for applications, 
as we shall see when we consider explicit models 
for illustrating  different types of bifurcation phenomena associated with NHIMs and reaction dynamics.

It is sufficient for our purpose to take phase space to be 
$\mathbb{R}^{2n}$, which is a $C^{\infty}$ differentiable manifold. 
For inner products and norms on this space we will take the usual Euclidean structures. 
Phase space is taken to be even dimensional since we will be considering canonical 
Hamiltonian systems, i.e. we consider equations of the following form:
\begin{subequations}
\label{eq:ham}
\begin{align}
\dot{q}_i & =  +\frac{\partial H}{\partial p_i} (q, p), \quad i=1, \ldots , n, \\ 
\dot{p}_i & =  -\frac{\partial H}{\partial q_i} (q, p), \quad i=1, \ldots , n,
\end{align}
\end{subequations}
for some function $H(q_1, \ldots, q_n, p_1, \ldots , p_n) \equiv H(q, p)$. 
We require the Hamiltonian $H(q, p)$ to be at 
least $C^{k+1}$ since we will assume the flow that it generates 
is at least $C^k$. In practice, this is not a restriction since the 
explicit Hamiltonians that we consider will be infinitely differentiable. 
However, in the theoretical description of NHIMs, it is important to explicitly  
specify the degree of differentiability of all geometrical objects for reasons 
that will become apparent as we develop the discussion of NHIMs further. 
Note that in this work we will have no need to consider $\mathbb{R}^{2n}$ as a symplectic vector space, 
or to investigate the associated  consequences of symplecticity; this is likely to be a very 
fruitful avenue for future research. The reason is the following. 
Roughly, in Hamiltonian dynamical systems stability properties are 
divided into `elliptic' and `hyperbolic', where bifurcation {\em may} occur 
when passing through the boundary between stability types. 
Results describing `elliptic stability' (e.g. the 
Kolmogorov-Arnold-Moser (KAM) and Nekhoroshev theorems) rely heavily on the Hamiltonian  
structure for their proof. `Hyperbolic stability' results, on the other hand, 
do not generally rely on the Hamiltonian structure, and that is 
true for the development of NHIMs. 
It would nevertheless be interesting to know what restrictions Hamiltonian structure 
places on the structure of NHIMs. 
For example, in ref.\ \onlinecite{bolotin2000remarks} 
it is proven that invariant tori in canonical, time independent, Hamiltonian systems {\em cannot} 
be normally hyperbolic.

We suppose that  $\Phi^t : \mathbb{R}^{2n} \times \mathbb{R} \rightarrow  \mathbb{R}^{2n}$ 
is a $C^k$ smooth flow defined  on $\mathbb{R}^{2n}$ generated by \eqref{eq:ham}. 
We will also need assume the flow to exist for all time, i.e. $t \in \mathbb{R}$.  
As phase space is not compact, we will need 
to explicitly discuss the issue of existence of the flow for all time in our examples.

We now state the definition of a NHIM sufficient for our needs.

\begin{definition}[NHIM]  An $\ell$ dimensional submanifold, $\Lambda$  of  $\mathbb{R}^{2n}$  
($\ell < 2n$) is said to be a normally hyperbolic invariant manifold for $\Phi^t$ if $\Lambda$ 
is invariant under $\Phi^t$ and there exists a splitting of the tangent bundle of  
$ T \mathbb{R}^{2n}  $ into sub-bundles:

\begin{equation}
T \mathbb{R}^{2n} = E^u \oplus E^s  \oplus T \Lambda,
\label{eq:splitting}
\end{equation}

\noindent
having the following properties:

\begin{description}

\item[Invariance.] The sub-bundles  are invariant under $D \Phi^t$, for all $ x \in \Lambda$,  
$t \in \mathbb{R}$, i.e.,
\begin{subequations}
\begin{align}
v \in E^s_x  & \Rightarrow    D \Phi^t (x) (v)  
\subset  E^s_{\Phi^t(x)}\quad \mbox{for all} \quad x \in \Lambda, \quad   t \in \mathbb{R},  \label{inv_s}\\
v \in E^u_x  & \Rightarrow   D \Phi^t (x) (v)  
\subset  E^u_{\Phi^t(x)}  \quad \mbox{for all}  \quad  x \in \Lambda, \quad   t \in \mathbb{R}, \label{inv_u}\\
v \in T \Lambda_x  & \Rightarrow   D \Phi^t (x) (v)  
\subset  T \Lambda_{\Phi^t(x)}\quad \mbox{for all}  \quad  x \in \Lambda, \quad   t \in \mathbb{R} \label{inv_t}
\end{align}
\end{subequations}

\item[Growth Rates.] There exists a constant $C > 0$ and rates 
$0  \le \beta < \alpha$ such that for all $x \in \Lambda$ we have:
\begin{subequations}
\begin{align}
v \in E^s_x  & \Rightarrow   
\parallel D \Phi^t (x) (v) \parallel  \le C e^{- \alpha t} 
\parallel v \parallel \quad \mbox{for all} \quad  t \ge 0,  \label{growth-s} \\
v \in E^u_x  & \Rightarrow 
\parallel D \Phi^t (x) (v) \parallel  \le C e^{\alpha t} 
\parallel v \parallel \quad \mbox{for all}  \quad t \le 0,  \label{growth-u}\\
v \in T_x \Lambda   & \Rightarrow 
\parallel D \Phi^t (x) (v) \parallel  \le C e^{\beta  \vert t \vert } 
\parallel v \parallel \quad \mbox{for all}  \quad t \in \mathbb{R}, \label{growth-t}
\end{align}
\end{subequations}

\end{description}

\label{def:NHIM}
\end{definition}

We now provide some additional background related to the concepts in this definition.

\begin{description}

\item[Invariance.] Invariance of $\Lambda$ under $\Phi^t$ means that 
for any $x \in \Lambda$, $\Phi^t (x) \in \Lambda$ for all $t \in \mathbb{R}$.  
We have {\em not} required $\Lambda$ to be boundaryless.  
Making the assumption that $\Lambda$ has no boundary is usual when 
considering invariance of $\Lambda$, since  otherwise one would need to make some 
sort of assumption preventing trajectories from leaving $\Lambda$ by crossing the boundary 
of $\Lambda$. We have {\em not} required $\Lambda$ to have no boundary 
since that is not natural for the examples we will consider. 
Since we will be considering only autonomous Hamiltonian systems 
the fact that trajectories are prevented from leaving by crossing the boundary of $\Lambda$  
is ensured by constancy of the Hamiltonian function, i.e. energy conservation. 

\item[Rates.] Note the rates $0  \le  \beta < \alpha$. These, 
along with \eqref{growth-s}, \eqref{growth-u}, and \eqref{growth-t} encapsulate the 
idea of {\em normal} hyperbolicity, in the sense that the rate of growth or decay  
of tangent vectors  {\em normal} to $\Lambda$ under the linearized dynamics dominate 
the rate of growth or decay of vectors tangent to $\Lambda$ under the linearized dynamics. 
Note that vectors tangent to $\Lambda$ can still exhibit exponential growth or decay, 
but the {\em linearized} rate must be smaller than that normal to $\Lambda$.

\item[Energy Conservation.]  The level set of the  Hamiltonian, 
\begin{equation}
\Sigma_E = \left\{ (q, p) \in \mathbb{R}^{2n} \vert H(q, p) = E \right\},
\label{eq:energy}
\end{equation}
is a $2n-1$ dimensional surface (except at possible singular points of the surface)  
that is invariant under the flow generated by Hamilton's equations. 
If it is compact and boundaryless, then the flow exists for all time. 
Generically \cite{footnote1} 
the $m$ dimensional manifold $\Lambda$ will intersect  \eqref{eq:energy} in a $\ell-1$  
dimensional manifold, which we refer to as $\Lambda_E$. If $\Lambda$ is normally hyperbolic, 
then $\Lambda_E \equiv \Lambda \cap \Sigma_E $ is also  normally hyperbolic.  
Thus, for Hamiltonian systems we will be considering both iso-energetic and non iso-energetic NHIMs.

\item[Dimensionality.] We will assume that:
\begin{equation}
{\rm dim} \, E^u_x = {\rm dim} \, E^s_x =m,
\label{dim-su}
\end{equation}
and it follows from the conditions of Definition \ref{def:NHIM} that $T_x \Lambda$, 
$E^u_x$ and $E^s_x$  vary continuously with respect to $x$, and therefore the dimensions 
of  $T_x \Lambda$,  $E^u_x$ and $E^s_x$  are constant with respect to $x$. We also have:
\begin{equation} 
{\rm dim} \, T_x \Lambda = \ell.
\label{dim-t}
\end{equation}
It follows from \eqref{eq:splitting}  that:
\begin{equation}
2n = 2m + \ell.
\label{dim_count}
\end{equation}
Hence, $\ell$ is even.

\item[The Need for Bundles: Linearization Characteristics.]  
The `bundle' concept and terminology, while perhaps unfamiliar, 
is necessary in our setting since we are considering the linearization about an invariant  manifold, 
which is generally filled with different orbits 
(as opposed to  the more familiar situation of linearizing about a single orbit, such as an equilibrium point  
or periodic orbit). 
Consider the unstable bundle, for example.  Invariance  
of this bundle under the linearized flow means that 
$v \in E^u_x   \Rightarrow    D \Phi^t (x) (v)  \subset  
E^u_{\Phi^t(x)}\quad \mbox{for all} \quad x \in \Lambda, \quad   t \in \mathbb{R}$.  
This relation provides an unstable growth direction all along the orbit through 
$x$, i.e. all along $\Phi^t (x), t \in \mathbb{R}$, and enables us 
to quantify  the unstable growth along a given orbit.
The  (disjoint) union of all of the unstable subspaces over 
all points in $\Lambda$ gives the unstable bundle over $\Lambda$. 
Similar consideration hold for the  stable bundle over $\Lambda$ and 
the tangent bundle of $\Lambda$. 
More details can be found in the general references for NHIMs given earlier.

\end{description}

\paragraph{Existence of Stable and Unstable Manifolds.} As we have previously mentioned,  
the definition of NHIM given in Definition \ref{def:NHIM} specifies conditions 
on the {\em linearized} dynamics about $\Lambda$.  
These provide {\em sufficient} conditions for proving results about the nonlinear dynamics.  
In particular, the sets of points having the following properties:
\begin{subequations}
\begin{align}
W^s (\Lambda) &  =  \left\{ y \in \mathbb{R}^{2n} \, \vert \, d \left( \Phi^t (y, M \right) 
\le C_y e^{- \alpha t} \quad \mbox{ for all} \quad t \ge 0 \right\} \\
W^u (\Lambda) &  =   \left\{ y \in \mathbb{R}^{2n} \, \vert \, d \left( \Phi^t (y, M \right) 
\le C_y e^{ \alpha t} \quad \mbox{ for all} \quad t \le 0  \right\}
\end{align}
\end{subequations}
can be shown to be $C^{s-1}$ manifolds (for some constant $C_y >0$), 
where $s < {\rm min} \left\{k, \frac{\alpha}{\beta} \right\}$, where, 
we recall, $k$ is the degree of differentiability of the flow.  
Moreover, under the conditions of Definition \ref{def:NHIM}, $\Lambda$  
can be shown to be a $C^s$ invariant manifold. 
However, we emphasise again that the conditions of 
Definition \ref{def:NHIM} provide {\em sufficient} conditions (in terms of linearized dynamics) 
for the smoothness properties of the manifolds.  
For example, in the models that we consider in Section \ref{sec:models}, $\Lambda$ 
is explicitly known and is infinitely differentiable, 
regardless of the value of the ratio $\frac{\alpha}{\beta}$.  
Similarly, the flow will be infinitely differentiable in the models that we consider in Section \ref{sec:models}.

\paragraph{Persistence Under Perturbation. } 
One of the most important properties of  NHIMs is that they persist under $C^1$ perturbations. 
It is important to clearly define the term `perturbation'. 
For example,  we are considering time independent  Hamiltonian systems. 
Time-dependent erturbations would not  satisfy the hypotheses of the perturbation theorem, 
unless the  time-dependence could be recast in a way that the system became time independent, 
and the resulting system satisfies the hypotheses of the perturbation theorem. 
Time dependence that is periodic or quasi periodic can be treated in this way, while
more general time dependence results in a breakdown of compactness, leading to 
problems satisfying the hypotheses of the theorem.

\paragraph{Sufficient Conditions and `Breakdown of Normal Hyperbolicity'} Definition \ref{def:NHIM}  
is a  definition of a NHIM. It {\em cannot} be concluded that if the conditions of the
definition are not satisfied, such as might occur for certain parameter values  
in a parameterized system, that the NHIM undergoes a `bifurcation'. 
Loss of hyperbolicity is a {\em necessary, but not sufficient} condition for  
bifurcation to occur, but the nature of a bifurcation depends on nonlinearity, 
which is outside the standard requirements of normal hyperbolicity as stated in Definition \ref{def:NHIM}. 

\subsection{The General Form of the Models Under Consideration}
\label{sec:gen_models}

We now consider the general form of a model that encompasses 
all of the specific examples that we will study. 
We consider a two DoF Hamiltonian system of the following form:
\begin{equation}
H(q_1, p_1, q_2, p_2) = H_1 (q_1, p_1) + H_2 (q_2, p_2), \quad (q_1, p_1, q_2, p_2) \in \mathbb{R}^4.
\label{gen_model_1}
\end{equation}
with associated Hamiltonian vector field:
\begin{subequations}
\label{gen_model_2}
\begin{align}
\dot{q}_1 & =  \frac{\partial H_1}{\partial p_1} (q_1, p_1), \\ 
\dot{p}_1 & =  -\frac{\partial H_1}{\partial q_1} (q_1, p_1), \\
\dot{q}_2 & =  \frac{\partial H_2}{\partial p_2} (q_2, p_2), \\
\dot{p}_2 & =  -\frac{\partial H_2}{\partial q_2} (q_2, p_2).
\end{align}
\end{subequations}
Note the particularly simple form of eqs \eqref{gen_model_2}.  
It has the form of two separable one degree-of-freedom Hamiltonian systems 
(hence it is completely integrable).  The flow generated by \eqref{gen_model_2} has the general form:
\begin{equation}
\Phi^t (q_{10}, p_{10}, q_{20}, p_{20} ) = \left(q_1 ( t, q_{10}, p_{10}), p_1 (t, q_{10}, p_{10}), q_2 ( t, q_{20}, p_{20}), p_2 (t, q_{20}, p_{20})
\right).
\label{gen_model_3}
\end{equation}
The equation for  the (typically)  three-dimensional energy surface is given by:
\begin{equation}
\Sigma_E = \left\{
(q_1, p_1, q_2, p_2) \in \mathbb{R}^4 \, \vert \, H(q_1, p_1, q_2, p_2) = H_1 (q_1, p_1) + H_2 (q_2, p_2) = E \right\}.
\label{gen_model_4}
\end{equation}

We make the following assumption on the $q_2-p_2$ component of \eqref{gen_model_2}.

\begin{assumption}
At $q_2=p_2 =0$ the system:
\begin{subequations}
\label{gen_model_5}
\begin{align}
\dot{q}_2 & =  \frac{\partial H_2}{\partial p_2} (q_2, p_2), \\
\dot{p}_2 & =  -\frac{\partial H_2}{\partial q_2} (q_2, p_2).
\end{align}
\end{subequations}
has a hyperbolic equilibrium point. The (positive) eigenvalue of the matrix 
associated with the linearization  of \eqref{gen_model_5} about this equilibrium is $\alpha > 0$.
\label{hyp_assump}
\end{assumption}

Clearly, the set
\begin{equation}
\Lambda = \left\{ (q_1. p_1, q_2, p_2) \, \vert \, q_2 = p_2 =0 \right\},
\label{gen_model_6}
\end{equation}
is a two dimensional invariant manifold in the four dimensional phase space.  
The intersection of the 3D energy surface with this 2D  invariant manifold is given by:
\begin{equation}
\Lambda_E=\Lambda \cap \Sigma_E =  \left\{
(q_1, p_1, q_2 ,p_2) \in \mathbb{R}^4 \, \vert \, H(q_1, p_1, 0, 0) = H_1 (q_1, p_1) + H_2 (0, 0) = E \right\},
\label{gen_model_7}
\end{equation}
which is  (typically) a one dimensional level set of $H_1 (q_1, p_1) $; 
an isoenergetic invariant manifold.  
Now we will show that, under Assumption \ref{hyp_assump}, $\Lambda$ and $\Lambda_E$ 
are both normally hyperbolic invariant manifolds.  
We therefore need to show that Definition  \ref{def:NHIM} holds for $\Lambda$. 

We begin by computing the linearization of the flow of  \eqref{gen_model_3} 
about an arbitrary point on $\Lambda$, which is denoted by $x \equiv (q_1, p_1, 0, 0)$.  
First, we transform $H_2 (q_2, p_2)$ to a set of coordinates that facilitates the computations. 
It follows from Assumption \ref{hyp_assump} that there exists a 
linear, symplectic transformation of coordinates, $(q_2, p_2) \rightarrow (\bar{q}_2, \bar{p}_2)$, 
such that the quadratic part of $H_2$ is ''diagonal'' in these coordinates, i.e.
\begin{equation}
H_2  (\bar{q}_2, \bar{p}_2) = H_2 (0, 0) + \alpha \bar{p}_2\bar{q}_2  + H^3  (\bar{q}_2, \bar{p}_2),
\label{H2_diag}
\end{equation}
where $H^3  (\bar{q}_2, \bar{p}_2)$ is ${\cal O}(3)$ 
(note: there are no linear terms in  $H_2  (\bar{q}_2, \bar{p}_2)$ 
since $(q_2, p_2) = (\bar{q}_2, \bar{p}_2) = (0, 0)$ is an equilibrium point). 
In these coordinates we can write the Hamiltonian \eqref{gen_model_1} in the form:
\begin{equation}
H(q_1, p_1, \bar{q}_2, \bar{p}_2) = H_1 (q_1, p_1) + 
H_2 ( \bar{q}_2, \bar{p}_2), \quad (q_1, p_1, \bar{q}_2, \bar{p}_2) \in \mathbb{R}^4.
\label{gen_model_diag}
\end{equation}

\noindent
In these coordinates, the linearisation of the flow has the  the following  block diagonal form:
\begin{equation}
D \Phi^t (x) = \left(
\begin{array}{cc}
A & 0_{2 \times 2} \\
0_{2 \times 2} &
\begin{array}{cc}
e^{\alpha t} & 0 \\
0 & e^{-\alpha t}
\end{array}
\end{array}
\right),
\label{gen_model_8}
\end{equation}
where
\begin{equation}
A = \left(
\begin{array}{rr}
\frac{\partial^2 H_1}{\partial q_1 \partial p_1} (q_1, p_1) & \frac{\partial^2 H_1}{\partial p_1^2 } (q_1, p_1) \\
-\frac{\partial^2 H_1}{\partial q_1^2} (q_1, p_1) & -\frac{\partial^2 H_1}{\partial q_1 \partial p_1 } (q_1, p_1) 
\end{array}
\right),
\label{gen_model_9}
\end{equation}
$0_{2 \times 2}$ denotes the $2 \times 2$ matrix of zeros, 
and we assume that we have  transformed the $q_2-p_2$ coordinates of the lower  right hand  $2 \times 2$ block  so that the flow assumes the diagonal form, which follows from Assumption \ref{hyp_assump}.

A general tangent vector in $E^u_x$, where we use the shorthand notation $x=(q_1, p_1, 0, 0)$, has the form
\begin{equation}
v = \left(
\begin{array}{c}
0 \\
0 \\
1 \\
0
\end{array}
\right),
\label{gen_model_10}
\end{equation}
 Then we have:
\begin{equation}
D \Phi^t (x) (v) =
 \left(
\begin{array}{c}
0 \\
0 \\
e^{\alpha t} \\
0
\end{array}
\right)
\label{gen_model_11}
\end{equation}
which is a tangent vector in $E^u_{x'}$, where $x' = \Phi^t (x)$.  
Since this argument holds for any $x \in \Lambda$ and any $v \in E^u_x$ 
it follows that the bundle $E^u$ is invariant (i.e. \eqref{inv_u} holds). 
Moreover, it follows from \eqref{gen_model_11} that the growth rate condition 
\eqref{growth-u} holds for all vectors in $E^u$. 
A similar argument for  the invariance of $E^s$ and the growth rate of vectors in $E^s$ follows.

Now we turn our attention to the tangent bundle of $\Lambda$. A general vector in $T_x \Lambda$ has the form:
\begin{equation}
v = \left(
\begin{array}{c}
a \\
b \\
0 \\
0
\end{array}
\right).
\end{equation}
Then using \eqref{gen_model_8} we have:
\begin{equation}
D \Phi^t (x) (v) =
 \left(
\begin{array}{c}
a' \\
b' \\
0 \\
0
\end{array}
\right)
\end{equation}
which is in  $T_{x'} \Lambda$, where $x' = \Phi^t (x)$.  
Here, $a'$ and $b'$ are functions of $(q_1, p_1)$, 
but the explicit functional form is not important for this argument.  
Since this argument holds for any $x \in \Lambda$, $v \in T_{x'} \Lambda$, it 
follows that $T \Lambda$ is invariant. What we can {\em not} verify at this point 
is the growth rate condition \eqref{growth-t}. This 
will require explicit conditions on the flow on $\Lambda$. 
The flow on $\Lambda$ can be very general, but normal hyperbolicity 
will require that \eqref{growth-t} is satisfied by this flow. 
This condition will be considered in the specific examples that we analyse in Section \ref{sec:models}.

\subsection{Phase space volumes, gap times, and reaction rates}
\label{sec:volumes}

In this section we briefly review the concepts from classical reaction 
rate theory that are relevant to the present study. 
This section is adapted from the paper of Collins et al\cite{collins2012isomerization} 
where more background and details can be found.

Points in the $4$-dimensional system phase space $\calM  = \mathbb{R}^{4}$ are denoted
$\bz \equiv \left(p_1, p_2, q_1, q_2 \right) \equiv (\mathbf{p}, \mathbf{q} )\in \calM$.
The system Hamiltonian is denoted by  $H(\bz)$, and
the $3$ dimensional  energy surface at energy $E$, $H(\bz) = E$, is denoted $\Sigma_E \subset \calM$.
The corresponding microcanonical phase
space density is $\delta(E - H(\bz))$, and the associated density of states
for the complete energy surface at energy $E$ is

\begin{equation}
\rho(E) = \Int{\bz}{\calM}{} \delta(E -  H(\bz)).
\end{equation}

The disjoint regions of phase space (`reactant', $q_2 <0$, and `product', $q_2 > 0$)
separated by the phase 
space dividing surface $\text{DS}(E)$  are denoted $\calM_{\pm}$;
the region of phase space corresponding to $q_2>0$  will be denoted by
$\calM_{\text{+}}$,  and that corresponding $q_2 <0$  will be denoted by
$\calM_{\text{-}}$.

The microcanonical density of states for points in region $\calM_{\text{+}}$  is
\begin{equation}
\rho_{\text{+}}(E) = \Int{\bz}{\calM_{\text{+}}}{} \delta(E - H(\bz))
\end{equation}
with a corresponding expression for the density of states $\rho_{\text{-}}(E)$ in $\calM_-$.
Since  the flow is everywhere transverse to $\text{DS}_{\pm}(E)$,
those phase points  in the  region $\calM_{\text{+}}$
that lie on  trajectories that cross $\text{DS}_{\pm}(E)$ can be  specified uniquely by coordinates $( \widetilde{p}, \widetilde{q}, \psi)$,
where $(\widetilde{p}, \widetilde{q}) \in \text{DS}_{\text{+}}(E)$ is a point on
$\text{DS}_{\text{+}}(E)$, specified by $2$
coordinates $(\widetilde{p}, \widetilde{q})$, and
$\psi$ is a time variable.
The point $\bz(\widetilde{p}, \widetilde{q}, \psi)$ is reached by propagating the
initial condition $(\widetilde{p}, \widetilde{q}) \in \text{DS}_{\text{+}}(E)$ forward for time $\psi$
\cite{thiele1962comparison,thiele1963comparison,ezra:164118}.
As all initial conditions on $\text{DS}_{\text{+}}(E)$
(apart from a set of trajectories of measure zero lying on stable manifolds)
will leave the region $\calM_{\text{+}}$ in finite time by crossing $\text{DS}_{\text{-}}(E)$, for each
$(\widetilde{p}, \widetilde{q}) \in \text{DS}_{\text{+}}(E)$, we can define the \emph{gap time}
$s = s(\widetilde{p}, \widetilde{q})$, which is the
time it takes for the trajectory to traverse the  region $\calM_{\text{+}}$ before entering the region $\calM_{\text{-}}$.
That is, $\bz(\widetilde{p}, \widetilde{q}, \psi = s(\widetilde{p}, \widetilde{q})) \in \text{DS}_{\text{-}}(E)$.
For the phase point $\bz(\widetilde{p}, \widetilde{q}, \psi)$, we therefore have
$0 \leq \psi \leq s(\widetilde{p}, \widetilde{q})$.

The coordinate transformation $\bz \to (E, \psi, \widetilde{p}, \widetilde{q})$ is canonical
\cite{arnol1989mathematical,thiele1962comparison,binney1985structure,meyer:3147}, 
so that the phase space volume element is
\begin{equation}
\label{coord_1}
\rmd^{4} \bz = \rmd E \, \rmd \psi  \, \rmd \sigma
\end{equation}
with $\rmd \sigma \equiv \rmd \widetilde{p} \, \rmd \widetilde{q}$
an element of $2$ dimensional area on  DS(E).

The magnitude $\phi(E)$ of the flux through dividing surface
$\text{DS}_{\text{+}}(E)$ at energy $E$ (`directional flux') is given by
\begin{equation}
\label{flux_1}
\phi(E) = \left\vert\Int{\sigma}{\text{DS}_{\text{+}}(E)}{} \right\vert,
\end{equation}
where the element of area $\rmd \sigma$ is precisely the restriction to DS(E) of the
appropriate flux $2$-form $\omega$ corresponding to the Hamiltonian vector field
associated with $H(\bz)$ \cite{toller1985theory,mackay1990flux,gillilan1990invariant,waalkens2004direct}.
The reactant phase space volume occupied by points initiated on the dividing surface
with energies between $E$ and $E + \rmd E$ is therefore
\cite{thiele1962comparison,brumer1980time,pollak1981classical,binney1985structure,meyer:3147,waalkens2005efficient,waalkens2005formula,ezra:164118}

\begin{subequations}
\label{vol_1}
\begin{align}
\rmd E \Int{\sigma}{\text{DS}_{\text{+}}(E)}{} \Int{\psi}{0}{s}
& = \rmd E \Int{\sigma}{\text{DS}_{\text{+}}(E)}{}  s \\
&= \rmd E \,\, \phi(E) \, \sbar
\end{align}
\end{subequations}
where the \emph{mean gap time} $\sbar$ is defined as
\begin{equation}
\sbar = \frac{1}{\phi(E)} \, \Int{\sigma}{\text{DS}_{\text{+}}(E)}{}  s
\end{equation}
and is a function of energy $E$.

\subsubsection{Gap time and reactant lifetime distributions}
\label{subsec:gaps}

The \emph{gap time distribution}, $\calP(s; E)$ is of central interest in
unimolecular kinetics \cite{slater:1256,thiele1962comparison}: the probability
that a phase point on $\text{DS}_{\text{+}}(E)$ at energy $E$ has a gap time between
$s$ and $s +\rmd s$ is equal to $\calP(s; E) \rmd s$.
An important idealized gap distribution is the random, exponential distribution
\begin{equation}
\label{exp_1}
\calP(s; E) = k(E) \, e^{-k(E) s}
\end{equation}
characterized by a single decay constant $k$ (where $k$ depends on energy $E$),
with corresponding mean gap time $\sbar = k^{-1}$.
An exponential distribution of gap times is usually  taken to be
a necessary condition for `statistical' behavior
in unimolecular reactions 
\cite{slater:1256,slater1959theory,thiele1962comparison,Dumont1986,carpenter2003nonexponential}.

The lifetime (time to cross the dividing surface $\text{DS}_{\text{-}}(E)$)
of phase point $\bz(\widetilde{p}, \widetilde{q}, \psi)$ is $t = s(\widetilde{p}, \widetilde{q}) - \psi$, and
the corresponding  (normalized)
reactant lifetime distribution function $\bbP(t; E)$ at energy $E$ is
\cite{slater:1256,slater1959theory,thiele1962comparison,bunker1962monte,bunker1964monte,bunker1973non,Dumont1986}
\begin{subequations}
\label{life_1}
\begin{align}
\label{life_1a}
\bbP(t; E) &= -\frac{\rmd}{\rmd t'}\; \text{Prob}(t \geq t'; E) \Big\vert_{t'=t} \\
\label{life_1b}
&= \frac{1}{\sbar} \, \Int{s}{t}{+\infty} \calP(s; E)
\end{align}
\end{subequations}
where the fraction of interesting (reactive) phase points having lifetimes between $t$ and $t + \rmd t$ is
$\bbP(t; E) \rmd t$.  It is often useful to work with the unnormalized lifetime distribution $F$,
where $F(t; E) \equiv  \sbar \, \bbP(t; E)$.

Equation \eqref{life_1a} gives the general relation between the lifetime distribution and the
fraction of trajectories having lifetimes greater than a certain value for arbitrary ensembles
\cite{bunker1962monte,bunker1964monte,bunker1973non}.
Note that an exponential gap time distribution \eqref{exp_1}
implies that the reactant lifetime
distribution $\bbP(t; E)$ is also exponential
\cite{slater:1256,slater1959theory,thiele1962comparison,bunker1962monte,bunker1964monte,bunker1973non}; both gap and lifetime distributions
for realistic molecular potentials have
been of great interest since the earliest days of trajectory simulations of
unimolecular decay, and many examples of non-exponential lifetime distributions
have been found
\cite{thiele1963comparison,bunker1962monte,bunker1964monte,bunker1966theory,bunker1968monte,bunker1973non,Hase76,grebenshchikov2003state,
lourderaj2009theoretical}.

\newpage

\section{Model Dynamics: Dividing Surfaces, Flux, Breakdown of Normal Hyperbolicity, and Gap Time Distributions}

In this section we describe our two model systems. Each model has NHIMs, 
and we describe the normal hyperbolicity properties of the NHIMs as a function of energy 
and other parameters. We then compute the flux through the associated DS as 
well as the gap time distribution, and we discuss the behaviour 
of these quantities in relation to the property of normal hyperbolicity.

\subsection{The Models}
\label{sec:models}

The two models to be studied 
are particular examples of the general class of models described in Section \ref{sec:gen_models}.

\subsubsection{Example 1. Uncoupled Simple Pendulum and Symmetric Double-Well}
\label{ex_1}

Our first example will consist of an integrable system whose Hamiltonian 
is the sum of the Hamiltonian of a simple pendulum and the Hamiltonian of a symmetric double-well:
\begin{equation}
H = \frac{p_1^2}{2} - \alpha_1^2 \cos q_1 + \frac{p_2^2}{2} - \frac{\alpha_2}{2}q_2^2 +  \frac{1}{4}q_2^4=H_1(q_1,p_1)+H_2(q_2,p_2),
\label{Ham_ex1}
\end{equation}
with associated equations of motion:
\begin{subequations}
\label{Hamvf_ex1}
\begin{align}
\dot{q}_1 & =  \frac{\partial H}{\partial p_1} = p_1,  \\ 
\dot{p}_1 & =  -\frac{\partial H}{\partial q_1} = - \alpha_1^2 \sin q_1, \\ 
\dot{q}_2 & = \frac{\partial H}{\partial p_2}  = p_2,  \\ 
\dot{p}_2 & =  -\frac{\partial H}{\partial q_2} = \alpha_2 q_2 - q_2^3
\end{align}
\end{subequations}

The phase space structure for these two, uncoupled, one degree-of-freedom Hamiltonian systems 
is illustrated in  Fig. \ref{fig:ex1_1}.   
Since the trajectories of \eqref{Hamvf_ex1} can be solved analytically, for any initial condition, it can be seen explicitly that the flow 
generated  by \eqref{Hamvf_ex1}  exists for all time. 

Following \eqref{gen_model_6},  the two dimensional non-isoenergetic surface:
\begin{equation}
\Lambda \equiv \left\{ (q_1, p_1, q_2, p_2) \in \mathbb{R}^4 \mid q_2 = p_2 =0 \right\},
\label{ex1_im}
\end{equation}
is an invariant manifold for \eqref{Hamvf_ex2}.  
For $\alpha_2 >0$, tangent vectors normal to $\Lambda$ experience exponential 
growth and decay under the linearised dynamics. 
However, we cannot claim that it is a NHIM according to definition \ref{def:NHIM}. 
Whether or not this is in fact the case 
depends on the dynamics on $\Lambda$. 
Following \eqref{gen_model_7}, the intersection of the three dimensional 
energy surface \eqref{Ham_ex1} with \eqref{ex1_im} is given by:
\begin{equation}
\Lambda_E \equiv \left\{ (q_1, p_1, q_2, p_2) \in \mathbb{R}^4 \mid \, 
\frac{p_1^2}{2} - \alpha_1^2 \cos q_2 =E, \,  q_2 = p_2 =0 \right\}.
\label{ex1_im_E}
\end{equation}
Hence, the dynamics on $\Lambda$ is described by the dynamics of a (symmetric) 
two well potential, and  an orbit of the pendulum with energy $E$ defines $\Lambda_E$. We note that:
\begin{itemize}

\item $q_1 = p_1 = 0$ is an elliptic equilibrium point of the pendulum with energy $E = -\alpha_1^2$.

\item Surrounding the elliptic fixed point is a family of periodic orbits 
(`librations') whose energies increase monotonically from $E = -\alpha_1^2$ to $E = \alpha_1^2$.

\item $q_1 = \pi, \, p_1 =0$ is a saddle point with energy $E= \alpha_1^2$ 
that is connected by a pair of homoclinic orbits.

\item Outside the pair of homoclinic orbits are two families of periodic 
orbits (`rotations')  whose actions increase monotonically with energy, from $E = \alpha_1^2$.

\end{itemize}

The phase space structure associated with the simple pendulum on $\Lambda$ is shown 
in  Fig.~\ref{fig:ex1_1}.   We note that homoclinic orbits have zero frequency. 
Hence, for one DoF Hamiltonian systems, they are examples of the phenomenon of {\em resonance}. 
The presence of homoclinic orbits therefore allows us to examine the role of resonance in the breakdown of 
normal hyperbolicity explicitly. 

In our model we take `reaction' to correspond to a change of sign of the coordinate $q_2$. 
Hence, the energy of reacting trajectories in the $q_2=p_2$ coordinates 
must be greater than that of the saddle point at $q_2=p_2 =0$, i.e., 
the energy must be greater than zero. Hence, 
the  total energy must be such that the energy in the $q_2-p_2$ subsystem is greater than zero.

\subsubsection{Example 2. Two Uncoupled Symmetric Double-Wells}
\label{ex_2}

The second example consists of two uncoupled symmetric double-well Hamiltonians:
\begin{subequations}
\label{eq:ex2_1}
\begin{align}
H & =  \frac{p_1^2}{2} - \frac{\alpha_1}{2}q_1^2 + 
\frac{1}{4}q_1^4 + \frac{p_2^2}{2} - \frac{\alpha_2}{2}q_2^2 +  \frac{1}{4}q_2^4 \\
& =  H_1(q_1,p_1)+H_2(q_2,p_2),
\end{align}
\end{subequations}
with associated Hamiltonian vector field:
\begin{subequations}
\label{Hamvf_ex2}
\begin{align}
\dot{q}_1 & =  \frac{\partial H}{\partial p_1} =   p_1,  \\
\dot{p}_1 & =  -\frac{\partial H}{\partial q_1} = \alpha_1 q_1 - q_1^3,  \\
\dot{q}_2 & =  \frac{\partial H}{\partial p_2}  = p_2, \\
\dot{p}_2 & =  -\frac{\partial H}{\partial q_2} = \alpha_2 q_2 - q_2^3.
\end{align}
\end{subequations}
Again, the trajectories of \eqref{Hamvf_ex2} can be solved analytically for any initial condition, and the flow 
generated  by \eqref{Hamvf_ex2}  exists for all time.

The discussion of the phase space structure is similar to that for Example 1 above. Following \eqref{gen_model_6},  
the two dimensional non-isoenergetic surface:
\begin{equation}
\Lambda \equiv \left\{ (q_1, p_1, q_2, p_2) \in \mathbb{R}^4 \mid q_2 = p_2 =0 \right\}
\label{ex2_im}
\end{equation}
is an invariant manifold for \eqref{Hamvf_ex2}.  
For $\alpha_2 >0$, tangent vectors normal to $\Lambda$ experience exponential 
growth and decay under the linearised dynamics. 
Again, whether or not $\Lambda$ is a NHIM according to definition \ref{def:NHIM}
depends on the dynamics on $\Lambda$. 
Following \eqref{gen_model_7}, the intersection of the three dimensional 
energy surface \eqref{eq:ex2_1} with \eqref{ex2_im} is given by:
\begin{equation}
\Lambda_E \equiv \left\{ (q_1, p_1, q_2, p_2) \in \mathbb{R}^4 \mid \, 
\frac{p_1^2}{2} - \frac{\alpha_1}{2} q_1^2 + \frac{1}{4} q_1^4=E, \,  q_2 = p_2 =0 \right\}.
\label{ex2_im_E}
\end{equation}

Hence, the dynamics on $\Lambda$ is described by the dynamics of a symmetric two well potential, 
and  an orbit of the symmetric two well potential  with energy $E$ defines $\Lambda_E$. We note that:
\begin{itemize}

\item $q_1 = p_1 = 0$ is a saddle point of the symmetric two well potential with energy $E = 0$.

\item  $q_1 = \pm \sqrt{\alpha_1}, \, p_1 =0$ are elliptic equilibrium points with energy $-\frac{1}{4} \alpha_1^2$.

\item Surrounding each elliptic equilibrium point is a family of periodic orbits 
whose energies increase monotonically away from the equilibrium point. 
At $E =0$ the periodic orbits merge into a pair
of homoclinic orbits that connect the saddle point at the origin.

\item As the energy increases from zero there is a family of (symmetric)
periodic orbits that surround both potential wells whose actions increase monotonically with energy.

\end{itemize}

The phase space structure associated with the symmetric double 
well on  $\Lambda$ is shown in  Fig. \ref{fig:ex2_1}.   

As in the previous model, `reaction' is associated with a change of sign of
the $q_2$ coordinate. Hence, the energy of reacting trajectories 
in the $q_2-p_2$ coordinates must be greater than the energy of  
the saddle point at $q_2=p_2 =0$, i.e. the energy must be greater than zero. 
The  total energy must therefore be such that the energy in the $q_2-p_2$ component of the model is greater than zero.  

\subsection{Dividing Surfaces and Flux}
\label{subsec:DS}

For each of our model Hamiltonians   we can construct a dividing surface in phase space
having the (local) no-recrossing  property. 
The codimension one non-isoenergetic surface defined by $q_2=0$ (locally) 
divides the phase space into two regions: one associated with the regions defined by $q_2 >0$ and the other associated with the region defined by $q_2 <0$. 
The dividing surface restricted to a three dimensional fixed energy
surface $ H\left(q_1, p_1, q_2, p_2 \right) = H_1 \left(q_1, p_1 \right) + H_2 \left( q_2, p_2 \right)  = E$  
is given by:
\begin{equation}
DS(E) = \left\{\left(q_1, p_1, q, p_2 \right) \, \vert \, q_2=0, \, 
H = H_1 (q_1, p_1) + \frac{p_2^2}{2} =E \right\}.
\label{eq:DS}
\end{equation}
This dividing surface has two halves:
\begin{subequations}
\begin{equation}
DS_+(E) = \left\{\left(q_1, p_1, q_2, p_2 \right) \, \vert \, q_2=0, \, H = H_1 (q_1, p_1) + \frac{p_2^2}{2} =E , \, p_2 >0\right\}
\label{eq:DS+}
\end{equation}
and 
\begin{equation}
DS_-(E) = \left\{\left(q_1, p_1, q_2, p_2 \right)  \, \vert \, q_2=0, \, H = H_1 (q_1, p_1) + \frac{p_2^2}{2} \bar{p}_2^2 =E ,  \, p_2 < 0\right\}.
\label{eq:DS-}
\end{equation}
\end{subequations}
These two halves meet at an invariant manifold:
\begin{equation}
\Lambda_E = \left\{\left(q_1, p_1, q_2, p_2 \right) \, \vert \, q_2=0, \, H = H_1 (q_1, p_1) =E,  \, p_2 = 0\right\}.
\label{eq:DSint}
\end{equation}
The nature of this invariant manifold (especially the normal hyperbolicity property) 
depends on $E$, the dynamics on $\Lambda_E$, and a comparison of the  linearised growth 
rates normal and tangent to $E$, as we will see for our examples. 
In particular, we will examine the inter-relation between these characteristics. 

We now describe conditions under which  $DS_+(E)$ and $DS_-(E)$ are surfaces having 
the no (local) re-crossing property (and we will verify these conditions in our particular examples).
These surfaces are defined by $q_2=0$. Therefore points on
these surfaces leave  if:
\begin{equation}
\dot{q}_2 = p_2 \neq 0,
\label{eq:leave}
\end{equation}
and therefore it follows immediately from their definition 
that \eqref{eq:DS+} and \eqref{eq:DS-} have the (local) `no-recrossing' property.

We denote the directional flux across these hemispheres by $\phi_{\text{+}} (E)$ and $\phi_{\text{-}} (E)$,
respectively, and note that $\phi_{\text{+}} (E)+ \phi_{\text{-}}(E)=0$.
The magnitude of the flux is $\vert \phi_{\text{+}} (E)\vert =  \vert \phi_{\text{-}}(E)\vert \equiv \phi (E)$.
The magnitude of the flux and related quantities are central to the theory of
reaction rates, as we described in Section \ref{sec:volumes}.

\subsubsection{Example 1. Simple pendulum plus Symmetric Double-Well}

\paragraph{\underline{Dividing surface.}}

In Figure \ref{figDSex1} we plot the dividing surface (i.e. \eqref{eq:DS}) for 
three different energies: $E = 0.5$ (energy below the separatrices of the pendulum), 
$E = \alpha_1^2 = 0.8^2$ (energy equal to the energy of the separatrices of the pendulum), and 
$E=1$ (energy larger than the energy of the separatrices of the pendulum). 
We see that the DS undergoes a bifurcation as the energy passes through the energy of the separatrix. 
A central question is whether or not this bifurcation has any effect on quantities that 
are important for quantifying reaction dynamics, such as flux and gap times.
~~\\

\paragraph{\underline{Flux across the DS.}}
By Stokes theorem, the directional  flux across  half the DS, at energy $E$,  is the area 
enclosed by the invariant manifold $\Lambda_E$ on the ($q_1$, $p_1$) plane.
This area is just the integral $A=\oint p_1\: dq_1$ and is related to the action 
variable of the simple pendulum by $J=\frac{1}{2\pi}\oint p_1\: dq_1$ so that we
have the relation $A=2\pi J$. For the simple pendulum there are 
three different cases to consider for the calculation of the action integral: 
the librations, the separatrices, and the rotations. 
The computation of these integrals for the different cases is reviewed in the
appendix \ref{appendixpendulum}. Here we just note the results:
 \begin{subequations}
 \label{pendulumactions}
 \begin{align}
J_l (E) & =  \frac{8\alpha_1}{\pi} \left[ \mathcal{E}_1 (k_l) + k_l^2 \mathcal{K}_1(k_l) 
- \mathcal{K}_1(k_l) \right]
\label{librationaction} \\
J_s (E) & =  \frac{4\alpha_1}{\pi}
\label{separatrixaction} \\
J_r (E) & =  \frac{4\alpha_1}{k_r \pi} \mathcal{E}_1 (k_r)
\label{rotationaction}
\end{align}
\end{subequations}
where the subscripts $l$, $s$ and $r$ denote libration, separatrix and rotation, respectively.
$\mathcal{K}_1$ and $\mathcal{E}_1$ denote the complete elliptic integrals of 
the first and second kind, respectively and
$k_l$ and $k_r$ are the moduli of the complete elliptic integrals, and are functions of the energy $E$:
\begin{subequations}
\label{pendulumactions_2}
\begin{align}
k_l (E) & =  \sin \left[ - \frac{1}{2} \cos^{-1} \left( \frac{E}{\alpha_1^2} \right) \right]
\label{librationmodulus} \\
k_r (E) & =  \sqrt{ \frac{2\alpha_1^2}{E + \alpha_1^2} }
\label{rotationmodulus}
\end{align}
\end{subequations}

With these expressions in hand we obtain expressions for 
the directional  flux $\phi(E)$ across  half the DS: 
\begin{subequations}
\label{pendulumflux}
\begin{align}
\phi_l (E) & =  16\alpha_1 \left[ \mathcal{E}_1 (k_l) + k_l^2 \mathcal{K}_1(k_l) - \mathcal{K}_1(k_l) \right]
\label{librationflux} \\
\phi_s (E) & =  16\alpha_1
\label{separatrixflux} \\
\phi_r (E) & =  \frac{16\alpha_1}{k_r} \mathcal{E}_1 (k_r)
\label{rotationflux}
\end{align}
\end{subequations}
For the separatrix and the rotation cases an extra factor 2 appears because one must count the area enclosed by
the curves having positive and negative momentum. The graph of the flux $\phi(E)$ is shown
in Figure \ref{figfluxpendulum}.

\subsubsection{Example 2. Uncoupled Symmetric Double-Wells}

\paragraph{\underline{Dividing surface.}}

In Figure \ref{figDSex2} we plot the dividing surface (i.e., the
surface \eqref{eq:DS}) for three different energies: 
$E = -0.5$ (energy below the separatrices of the two well potential), $E = 0$ 
(the energy of the separatrices of the two well potential), 
and $E=1$ (energy larger than the energy of the separatrices of the two well potential). 
As for example 1,  the DS undergoes a bifurcation as the energy passes through the energy of the separatrix.
~~\\

\paragraph{\underline{Flux across the DS.}}
As for Example 1, the directional flux across half the DS is the 
area enclosed by the invariant manifold $\Lambda_E$ on the ($q_1$, $p_1$) plane.
This area is  the integral $A=\oint p_1 \: dq_1$ and is related to 
the action variable of the symmetric double-well by 
$A=2\pi J$. For the symmetric double-well there are three different cases for the calculation of the action integral depending on the
energy being less, greater or equal to the energy of the saddle point. 
The integrals for the different cases are (cf.\ appendix \ref{appendixdoublewell}):
\begin{subequations}
\label{dwactions}
\begin{align}
J_1 (E) & = \frac{b \delta}{3 \sqrt{2} \pi k_1^2} \left[ (2-k_1^2) 
\mathcal{E}_1 (k_1) + (2k_1^2-2) \mathcal{K}_1(k_1) \right]
\label{dw1action} \\
J_2 (E) & =  \frac{2\alpha_2^{3/2}}{3\pi}
\label{dw2action} \\
J_3 (E) & =  \frac{\sqrt{2\delta} b^2}{3k_3^2 \pi} 
\left[ (2k_3^2-1) \mathcal{E}_1 (k_3) + (1-k_3^2) \mathcal{K}_1(k_3) \right]
\label{dw3action}
\end{align}
\end{subequations}
where the subscript 1, 2 and 3 refers to the three different regimes of energy, 
that is, energy below, and above the energy of the saddle point,  respectively. 
In theses expressions $a$ and $b$ are
roots of the equation $p_1 = 0$ (\textit{cf.} Appendix \ref{appendixdoublewell}) and $\delta=b^2-a^2$. 
All these 3 quantities 
are functions of the energy.
As for the case of the pendulum, the moduli $k_1$ and $k_3$ of the elliptic integrals are functions 
of the energy:
\begin{subequations}
\label{dwmodulus}
\begin{align}
a (E) & =  \sqrt{ \alpha_1 - \sqrt{\alpha_1^2+4E}}
\label{aroot} \\
b (E) & =  \sqrt{ \alpha_1 + \sqrt{\alpha_1^2+4E}}
\label{broot} \\
k_1 (E) & =  \sqrt{\frac{\delta(E)}{b^2(E)}}
\label{dw1modulus} \\
k_3 (E) & =  \sqrt{\frac{b^2(E)}{\delta(E)}}.
\label{dw3modulus}
\end{align}
\end{subequations}
The directional flux $\phi(E)$ across half  of the DS is then given by: 
\begin{subequations}
\label{dwflux}
\begin{align}
\phi_1 (E) & = \frac{4b \delta}{3 \sqrt{2} k_1^2} 
\left[ (2-k_1^2) \mathcal{E}_1 (k_1) + (2k_1^2-2) \mathcal{K}_1(k_1) \right]
\label{dw1flux} \\
\phi_2 (E) & =  \frac{8\alpha_2^{3/2}}{3}
\label{dw2flux} \\
\phi_3 (E) & =  \frac{2\sqrt{2\delta} b^2}{3k_3^2} 
\left[ (2k_3^2-1) \mathcal{E}_1 (k_3) + (1-k_3^2) \mathcal{K}_1(k_3) \right].
\label{dw3flux}
\end{align}
\end{subequations}
A plot of the flux $\phi(E)$ as a function of energy $E$ is shown in Figure \ref{figfluxex2}.

\subsection{Gap Time Distributions}
\label{subsec:gapdistrib}

In this subsection we  compute the gap time distribution of our two examples. Before
proceeding to the calculation of the gap time distribution, we need to determine the 
gap time for reactant trajectories.
In our two examples the determination of the gap times of reacting orbits is 
straightforward since the gap time is just half of the period of the
periodic orbits in the double-well degree-of-freedom  ($q_2$-$p_2$) 
for which the energy is greater than the energy of the saddle
point ($q_2=0$, $p_2=0$). The period of these periodic orbits as a function of the energy is given by:
\begin{equation}
T(E_2) = \frac{4\sqrt{2}}{\sqrt{\delta(E_2)}} \mathcal{K}(k_3(E_2)),
\label{periodPO}
\end{equation}
where $E_2$ stands for energy in the independent second degree-of-freedom, 
and the total energy of the two, uncoupled one degree-of-freedom 
systems is denoted by $E_T=E_1+E_2$.
The gap time as a function of the energy $E_2$ is then given by:
\begin{equation}
s(E_2) = \frac{2\sqrt{2}}{\sqrt{\delta(E_2)}} \mathcal{K}(k_3(E_2)),
\label{gaptime}
\end{equation}

For clarity here we will use the notations $I$ for the action associated with the first degree-of-freedom   
(pendulum for example one and
double-well for example 2) and $J$ for the action associated with the second degree-of-freedom.  
Because of the direct relation between
the  energies of the different degrees-of-freedom  and the actions associated with these 
degrees-of-freedom , at constant energy we have:
\begin{equation}
J =  J(E_T,I),
\label{actionJ}
\end{equation}
so that we can formally write the gap time as a function of the total energy and the action $I$: $s(E_2)=s(E_T,I)$.
Again at constant $E_T$, the number of trajectories having $s<\bar{s}=\bar{s}(E_T,\bar{I})$ is proportional to the
area enclosed by the curve $I=\bar{I}$ in the ($q_1$-$p_1$) plane, that is to say, it equals  $2\pi \bar{I}$. The
number of trajectories $n$ for which the gap time $s$ has the property 
that $\bar{s} \leq s \leq \bar{s}+ \Delta s$ is:
\begin{equation}
n =  2\pi \left[ (\bar{I}+\Delta I) - \bar{I} \right] = 2 \pi \frac{\partial I}{\partial s} \Delta s.
\label{numtraj}
\end{equation}
The gap time distribution is now given by:
\begin{subequations}
\label{gaptimedistrib}
\begin{align}
P(s;E_T) & =  \frac{n}{\Delta s} \\
& = 2 \pi \frac{\partial I}{\partial s} \\
& = \frac{2\pi}{  \left. \frac{\partial s}{\partial I} \right|_{E_T}}.
\end{align}
\end{subequations}
Figure \ref{figgaptimedistrib} shows the 
gap time distribution for both examples.

Figure \ref{figgaptimedistrib} shows that the gap time distributions for both examples 
possess a singularity. This singularity arises from the vanishing of the derivative
of the action $I$ with respect to the gap time $s$, which is equivalent to
the singularity of the derivative of this action
with respect to the energy $E_1$. This singularity appears because the action variable $I$ is a piecewise defined
function with respect to the energy $E_1$. Whereas the action $I$ is a continuous function with respect to  the energy $E_1$, its
derivative (the inverse of the associated frequency)
diverges at the separatrix (homoclinic orbits), which is exactly where  bifurcation occurs.
Interpreted geometrically, we see that 
just below the bifurcation the rate of growth of the area enclosed by the periodic orbits  in the 
($q_1$-$p_1$) plane increases to  infinity as we get closer to the homoclinic orbit. 
In the same way, just after the
bifurcation this growth rate decreases from infinity. 
The appearance of homoclinic orbits is manifested 
in the flux $\phi(E_T)$,  which 
is directly related to the action $I$, through the 
appearance of an inflexion point at the energy of the bifurcation.

\subsection{Loss of normal hyperbolicity and its consequences on the dynamics of reaction.}
\label{subsecnormalhyperbol}

In this section we will turn our attention to the question of normal hyperbolicity of 
the isoenergetic invariant manifold, which forms the boundary of 
the isoenergetic dividing surface. For both examples $\Lambda_E$  is 
either a periodic orbit or the union of homoclinic orbits and the saddle points that they connect. 
The two dimensional non-isoenergetic invariant manifold is defined by $q_2 = p_2 =0$, 
which is a hyperbolic saddle point with associated growth rate $\alpha_2$. 
For $\alpha_2 >0$ tangent vectors normal to $\Lambda$ experience exponential 
growth and decay under the linearised dynamics. Hence one way that normal hyperbolicity could 
be lost is for $\alpha_2$ to go from positive to negative. However, 
this possibility is not interesting from the point of view of reaction dynamics 
since this  would destroy the reactive (double well) nature of the problem. 
Hence, we will require $\alpha_2 >0$.

The more interesting situation is the violation of the growth 
condition \eqref{growth-t} in the definition of a NHIM (Definition \ref{def:NHIM}). 
When $\Lambda_E$ is a periodic orbit, this growth rate is zero, and $\Lambda_E$ is a NHIM. 
When $\Lambda_E$ is a union of homoclinic orbits and the saddle points that they 
connect, whether or not $\Lambda_E$ satisfies the growth rate conditions of Definition 
\ref{def:NHIM} depends on the nature of $\alpha_1$ and $\alpha_2$.  
In particular, for Example 1 $\pm \alpha_1$ are the eigenvalues associated 
with the saddle point  on the $q_1-p_1$ plane and for
Example 2 $\pm \sqrt{ \alpha_1}$ are the eigenvalues associated with the saddle point in the $q_1=p_1$ plane. 

We therefore have the following situations:

\begin{description}

\item[Example 1:] For $\alpha_1 >  \sqrt{\alpha_2}$ the growth rate conditions of Definition \ref{def:NHIM} are not satisfied, and $\Lambda_E$ is not a NHIM.

\item[Example 2:] For $\alpha_1  > \alpha_2$ the growth rate conditions of Definition \ref{def:NHIM} are not satisfied, and $\Lambda_E$ is not a NHIM.

\end{description} 

We now examine the consequences of this `loss of normal hyperbolicity'  in more  detail.

\subsubsection{Consequences of loss of normal hyperbolicity}
\label{sec:conseq}

We imagine the parameters $\alpha_1 >  \sqrt{\alpha_2}$  fixed for example 1
and  $\alpha_1  > \alpha_2$ for example 2, and we vary the energy in such a 
way that we pass through the energy value corresponding to the 
homoclinic orbits and saddle points. 
With $\alpha_1$ and $\alpha_2$ fixed as above, 
at this energy value the conditions 
of Definition \ref{def:NHIM} do not hold. 
This situation could be referred to as `loss of normal hyperbolicity'.  
We examine the implication of this loss of normal hyperbolicity 
for the quantities that we have computed for our two examples.

\begin{description}

\item[The Dividing Surface and the (Directional) Flux Through the Dividing Surface.]

The dividing surface, as a function of energy,  is shown in  Figure \ref{figDSex1} 
for example 1 and in Figure \ref{figDSex2} for example 2. 
While in both examples the geometry of the surface undergoes a qualitative change 
is we pass through the bifurcation, we are still able to define a dividing surface having the no recrossing
property as we pass through the bifurcation,  i.e., even as the NHIM experiences a loss of normal hyperbolicity
(as just defined).

The directional flux, as a function of energy,  
is shown in  Figure \ref{figfluxpendulum} for example 1 and 
in Figure \ref{figfluxex2} for example 2.
For both examples we see that the flux varies continuously 
as a function of the  energy, even as we pass through the bifurcation, i.e., 
even as the NHIM experiences a loss of normal hyperbolicity.

\item[The Gap Time Distribution.]

From Figure \ref{figgaptimedistrib} we have already noted 
note that the gap time distributions for both examples 
possess a singularity. This singularity is associated with the existence of
a homoclinic orbit in the DS, and is \emph{not} related to the 
loss of normal hyperbolicity of the NHIM.

\end{description}

\newpage

\section{Conclusions and Outlook}
\label{sec:concl}

In this paper we have studied the breakdown of normal hyperbolicity and 
its consequences for quantities related to reaction dynamics; 
in particular, the dividing surface, the flux through the dividing surface, 
and the gap time distribution. 

Our approach is to study these questions using simple, two degree-of-freedom Hamiltonian 
models for which calculations for the different geometrical and dynamical quantities can be carried out exactly. 
For our examples, we showed that resonances (homoclinic orbits)
within the normally hyperbolic invariant manifold may, 
or may not, lead to `loss of normal hyperbolicity'. 
Moreover, we showed that  for our examples the onset of such resonances 
results in a change in topology of the dividing surface, 
but it does not affect our ability to define a dividing surface (DS),
and that the flux through the DS varies continuously with energy, 
even as the energy is varied in such a way that normal hyperbolicity is lost. 
For both our examples we have shown that the gap time distribution  exhibits a signature
singularity at energies corresponding to emergence of a homoclinic orbit in the
DS, but these singularities are not
associated with loss of normal hyperbolicity.

\acknowledgments

FM, PC, and SW  acknowledge the support of the  Office of Naval Research 
(Grant No.~N00014-01-1-0769) and the Leverhulme Trust.

\clearpage

\appendix

\section{Action variables for the simple pendulum}
\label{appendixpendulum}

In this appendix we review the  compution of action variables for the integrable simple pendulum system.
The Hamiltonian of the simple pendulum is:
\begin{equation}
H =  \frac{p^2}{2} - \alpha_1^2 \cos q.
\label{App11}
\end{equation}

In order to determine the action variable for this system we have to consider three 
different cases: libration ($E<\alpha_1^2$), separatrix ($E=\alpha_1^2$), and rotation ($E>\alpha_1^2$).

\subsection{Libration}

The action for this case is given by:
\begin{equation}
J_l =  \frac{1}{2\pi} \oint p \: dq.
\label{App12}
\end{equation}
For this case the system doesn't have enough energy to cover the full range of the 
angle $q$ and there should be two turning points where the 
momentum vanishes, $p=0$. Let $q_0$ be the positive value of $q$ 
at the turning point which depends on the energy:
\begin{equation}
q_0 =  \cos^{-1} \left(-\frac{E}{\alpha_1^2} \right).
\label{App13}
\end{equation}

The momentum $p$ can be expressed as a function of $q_0$ and $q$:
\begin{equation}
p =  \sqrt{ 2\alpha_1^2 (\cos q - \cos q_0)}.
\label{App14}
\end{equation}
Substituting this expression into the eq.\ \eqref{App12} we obtain the integral:
\begin{equation}
J_l =  \frac{4\alpha_1}{\pi} \int_0^{q_0} \sqrt{\sin^2\left(\frac{q_0}{2}\right) - \sin^2 \left(\frac{q}{2} \right)} \: dq.
\label{App15}
\end{equation}
After an appropriate change of variable we can put the former integral into a form which makes it resemble an 
elliptic integral: 
\begin{equation}
J_l =  \frac{8\alpha_1k_l^2}{\pi} \int_0^{\pi/2} \frac{\cos^2 \phi}{\sqrt{1-k_l^2\sin^2 \phi}} d\phi,
\label{App16}
\end{equation}
where the modulus $k_l$ of the elliptic integral is given by $k_l=\sin\left(\frac{q_0}{2}\right)$.
This integral can be expressed in terms of complete elliptic integrals as follows:
\begin{equation}
J_l =  \frac{8\alpha_1}{\pi} \left[ \mathcal{E}_1(k_l) + k_l^2 \mathcal{K}_1 (k_l) -  \mathcal{K}_1 (k_l) \right],
\label{App17}
\end{equation}
where $\mathcal{K}_1$ and $\mathcal{E}_1$ stand for the complete elliptic integrals of the first and second
kinds respectively \cite{hancock1917elliptic}.

\subsection{Rotation}

For this case the system has enough energy to cover the full range
of the angle $q$ and there is no point where the momentum vanishes.
The momentum is given by:
 \begin{equation}
p =  \sqrt{2(E+\alpha_1^2\cos q)}.
\label{App18}
\end{equation}
The corresponding action integral is:
 \begin{equation}
J_r =  \frac{1}{2\pi} \int_0^{2\pi} \sqrt{2(E+\alpha_1^2 \cos q)} \:dq.
\label{App19}
\end{equation}
Again using a change of variables we can transform this integral to:
 \begin{equation}
J_r =  \frac{2\alpha_1}{k_r\pi} \int_0^{\pi} \sqrt{1-k_r^2 \sin^2 \phi} \:d\phi,
\label{App110}
\end{equation}
where the modulus $k_r$ is given by $k_r=\sqrt{\frac{2\alpha_1^2}{E+\alpha_1^2}}$.
This integral is just twice the complete elliptic integral of the second kind, so
that the action for the rotation case is:
 \begin{equation}
J_r =  \frac{4\alpha_1}{k_r\pi} \mathcal{E}_1 (k_r).
\label{App111}
\end{equation}

\subsection{The separatrix case}

For the separatrix case we can repeat the same kind of calculations as for the rotation case.
Noting that for this case we have $k_r=1$ and $\mathcal{E}_1(1)=1$ the action
for the separatrix case is just:
 \begin{equation}
J_s =  \frac{4\alpha_1}{\pi}.
\label{App112}
\end{equation}

\section{Action variables for the symmetric double-well}
\label{appendixdoublewell}

In this appendix we compute action variables for the integrable symmetric double-well system.
The Hamiltonian of the system is:
\begin{equation}
H =  \frac{p^2}{2} - \alpha_2 \frac{q^2}{2} + \frac{q^4}{4}.
\label{App21}
\end{equation}

In order to determine the action variable of this system we have to consider three different cases 
distinguished by the value of the energy $E$: $E<0$,
the separatrix case where $E=0$, and $E>0$.

\subsection{Case $E<0$.}

For this case the energy is below the saddle point energy, $E=0$. 
Trajectories are confined to one of the two wells.
Because the double-well is symmetric the action variable is the same for the motions in
either well.From eq.\ \eqref{App21} we can write the momentum $p$ as a function of the energy and $q$:
\begin{equation}
p =  \sqrt{ 2E + \alpha_2  q^2 - \frac{q^4}{2} }.
\label{App22}
\end{equation}
The associated action integral is:
\begin{equation}
J_1 =  \frac{1}{2\pi} \oint p \: dq = \frac{1}{2\pi} 2 \int_a^b \sqrt{ 2E + \alpha_2  q^2 - \frac{q^4}{2} } \: dq,
\label{App23}
\end{equation}
where $a$ and $b$ are the two roots of the equation $p=0$ for the well situated situated on the side where $q>0$
for example and with $a<b$.

Using a change of variable we can rearrange the result into the form:
\begin{equation}
J_1 =  \frac{b\delta k_1^2}{\sqrt{2}\pi} \int_0^{\pi/2} \frac{\sin^2 \theta \cos^2 \theta}{\sqrt{1-k_1^2 \sin^2 \theta} } \: d\theta,
\label{App24}
\end{equation}
where $\delta=b^2-a^2$ and $k_1=\sqrt{\frac{\delta}{b^2}}$. 
The latter integral can be evaluated by parts and we
finally obtain:
\begin{equation}
J_1 =  \frac{b\delta}{3\sqrt{2}\pi k_1^2} \left[ (2-k_1^2) 
\mathcal{E}_1 (k_1) + (2 k_1^2 - 2) \mathcal{K}_1 (k_1) \right].
\label{App25}
\end{equation}

\subsection{Case $E>0$.}

For this case the energy is above the saddle point energy. In this case the equation $p = 0$ has only two roots 
situated symmetrically with respect to the coordinate origin $q=0$.
Denoting the two roots $a$ and $b$, $a < b$, the action integral is:
\begin{equation}
J_3 =  \frac{1}{2\pi} \oint p \:dq = \frac{4}{2\pi} \int_0^b p \:dq,
\label{App26}
\end{equation}
Using an appropriate change of variable
we can transform this integral to the form:
\begin{equation}
J_3 = \frac{\sqrt{2\delta} b^2}{\pi} \int_0^{\pi/2} \sqrt{1-k_3^3 \sin^2 \theta} \sin^2 \theta \: d\theta,
\label{App27}
\end{equation}
where $\delta=b^2-a^2$ and $k_3=\sqrt{\frac{b^2}{\delta}}$. 
Inetgrals of this type are tabulated (see for example ref.\
\onlinecite{hancock1917elliptic}) and one gets the result:
\begin{equation}
J_3 = \frac{\sqrt{2\delta} b^2}{3\pi k_3^2} 
\left[ (2 k_3^2-1) \mathcal{E}_1 (k_3) + (1-k_3^2) \mathcal{K}_1 (k_3) \right].
\label{App28}
\end{equation}

\subsection{Case $E=0$.}

The action in this case is obtained by setting $a=0$, $b=\sqrt{2\alpha_2}$, $\delta=b^2$
and $k_3=1$ in the previous case, and noting also that instead of having
a factor 4 in eq.\ \eqref{App26} we have here a factor 2.
Taking into account the fact that $\mathcal{E}_1(1)=1$ we get for the separatrix action:
\begin{equation}
J_2 = \frac{2\alpha_2^{3/2}}{3\pi}.
\label{App29}
\end{equation}

\newpage

\section*{Figure captions}

\begin{figure}[H]
\caption{Phase portrait for example 1 (cf.\ sec.\ \ref{ex_1}).  
The phase space is the Cartesian product (denoted by ``{\sf X}'') of the phase spaces for the uncoupled subsystems.}
\label{fig:ex1_1}
\end{figure}

\begin{figure}[H]
\caption{Phase portrait for example 2 (cf.\ sec.\ \ref{ex_2}).
The phase space is the Cartesian product (denoted by ``{\sf X}'') of the phase spaces for the uncoupled subsystems.}
\label{fig:ex2_1}
\end{figure}

\begin{figure}[H]
\caption{
Dividing surface for example 1 (eq.\ \eqref{eq:DS}) for 
three different energies.  Parameter $\alpha_1=0.8$. (a) $E=0.5$. (b) $E=\alpha_1^2$. (c) $E=1.0$.}
\label{figDSex1}
\end{figure}

\begin{figure}[H]
\caption{Flux $\phi(E)$ across the DS for example 1 as a function of energy $E$ ($\alpha_1=0.8$).}
\label{figfluxpendulum}
\end{figure}

\begin{figure}[H]
\caption{Dividing surface for example 2 (eq.\ \eqref{eq:DS}) for three different energies.
Parameters $\alpha_1=\alpha_2=1$. (a) $E=-0.15$. (b) $E=0$. (c) $E=1.0$.}
\label{figDSex2}
\end{figure}

\begin{figure}[H]
\caption{Flux $\phi(E)$ across the DS for example 2 as a function of energy $E$ ($\alpha_1=\alpha_2=1$).}
\label{figfluxex2}
\end{figure}

\begin{figure}[H]
\caption{Gap time distributions. (a) Example 1, energy $E_T=5$. (b) Example  2, energy $E_T=5$.  
Note the existence of a singularity in the gap time distributions for both cases.}
\label{figgaptimedistrib}
\end{figure}



\newpage

\begin{center}
\includegraphics[width=10cm]{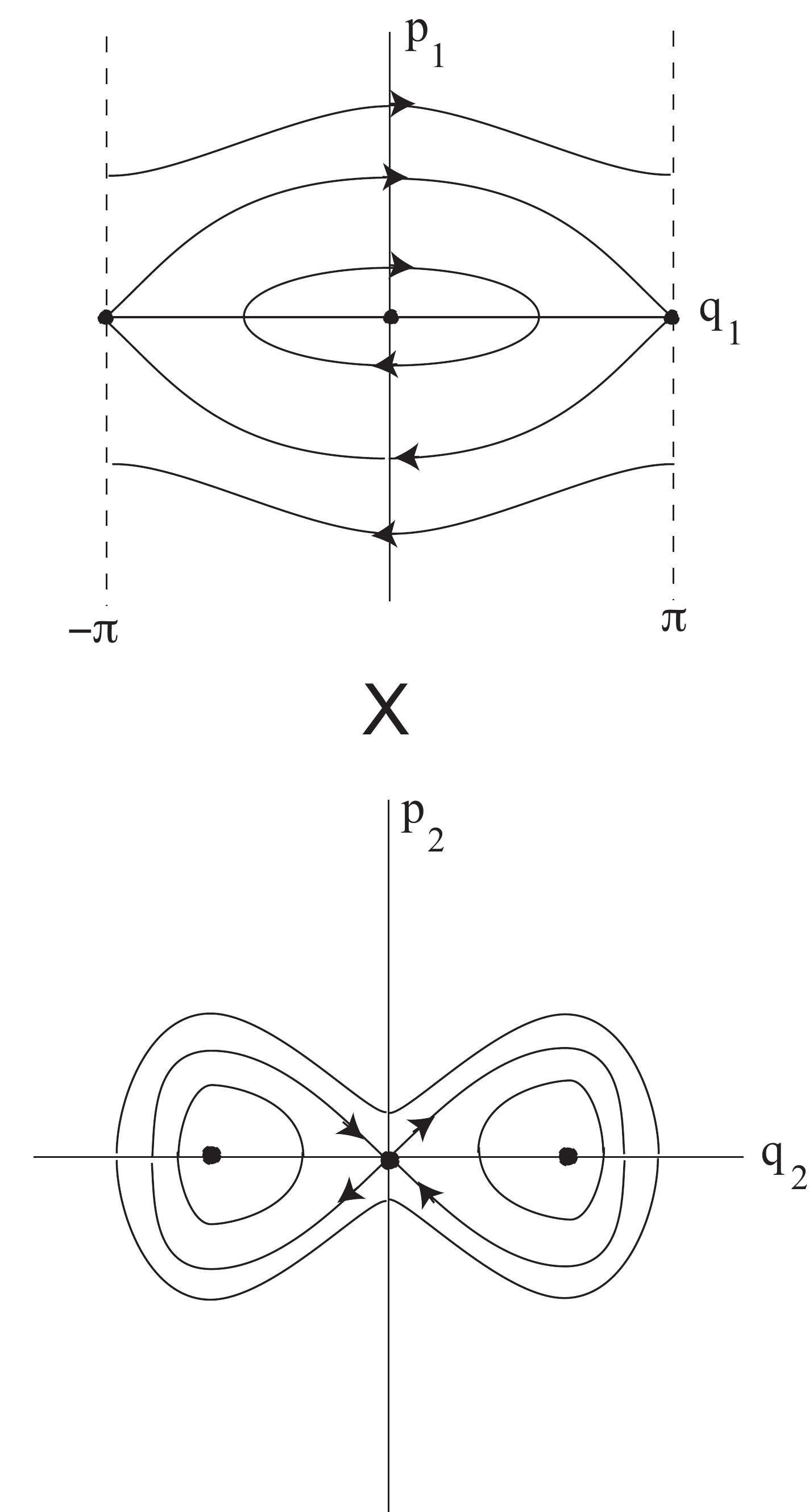}
\end{center}

\vspace*{1.5cm}
FIGURE 1
 
\newpage

\begin{center}
\includegraphics[width=9cm]{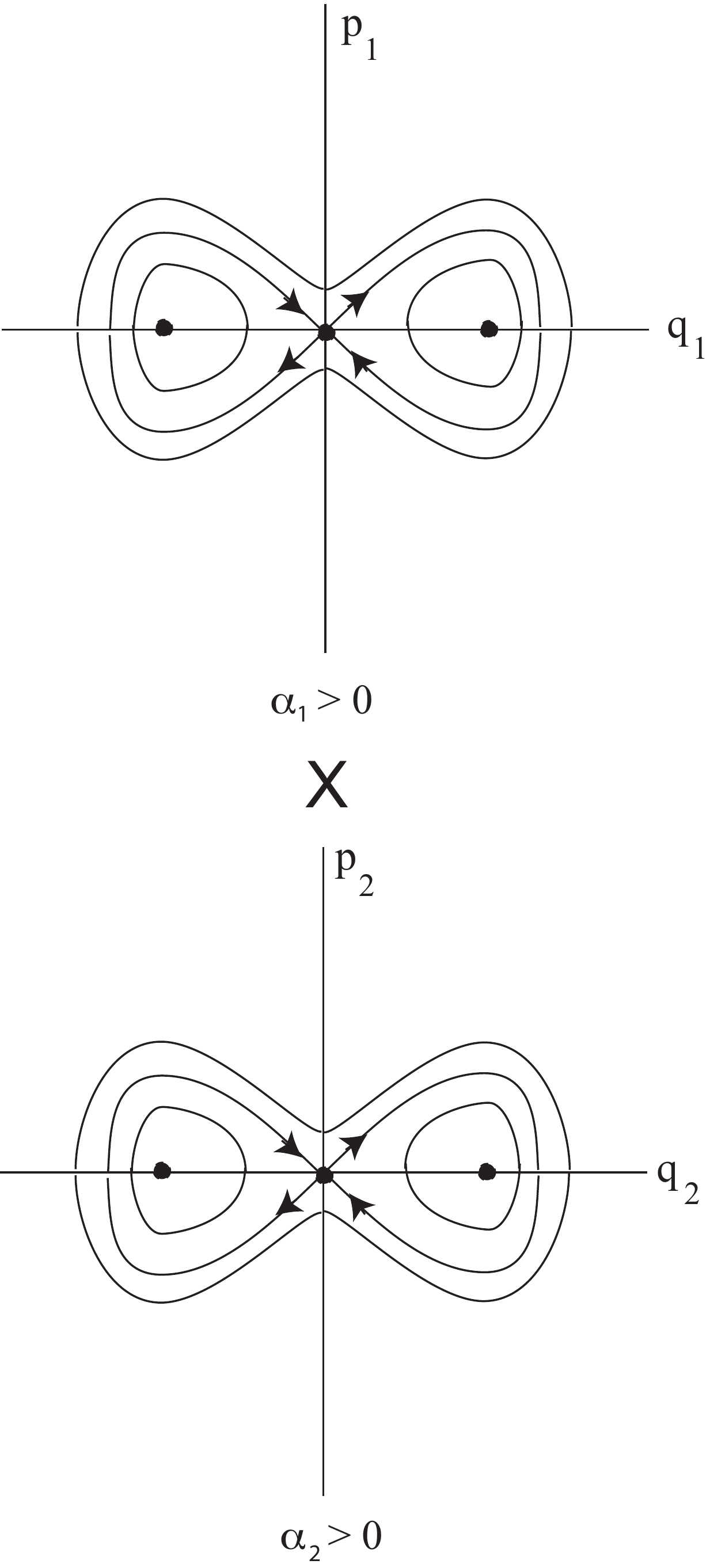}
\end{center}

\vspace*{1.5cm}
FIGURE 2
 
\newpage

\begin{center} 
\includegraphics[scale=0.3]{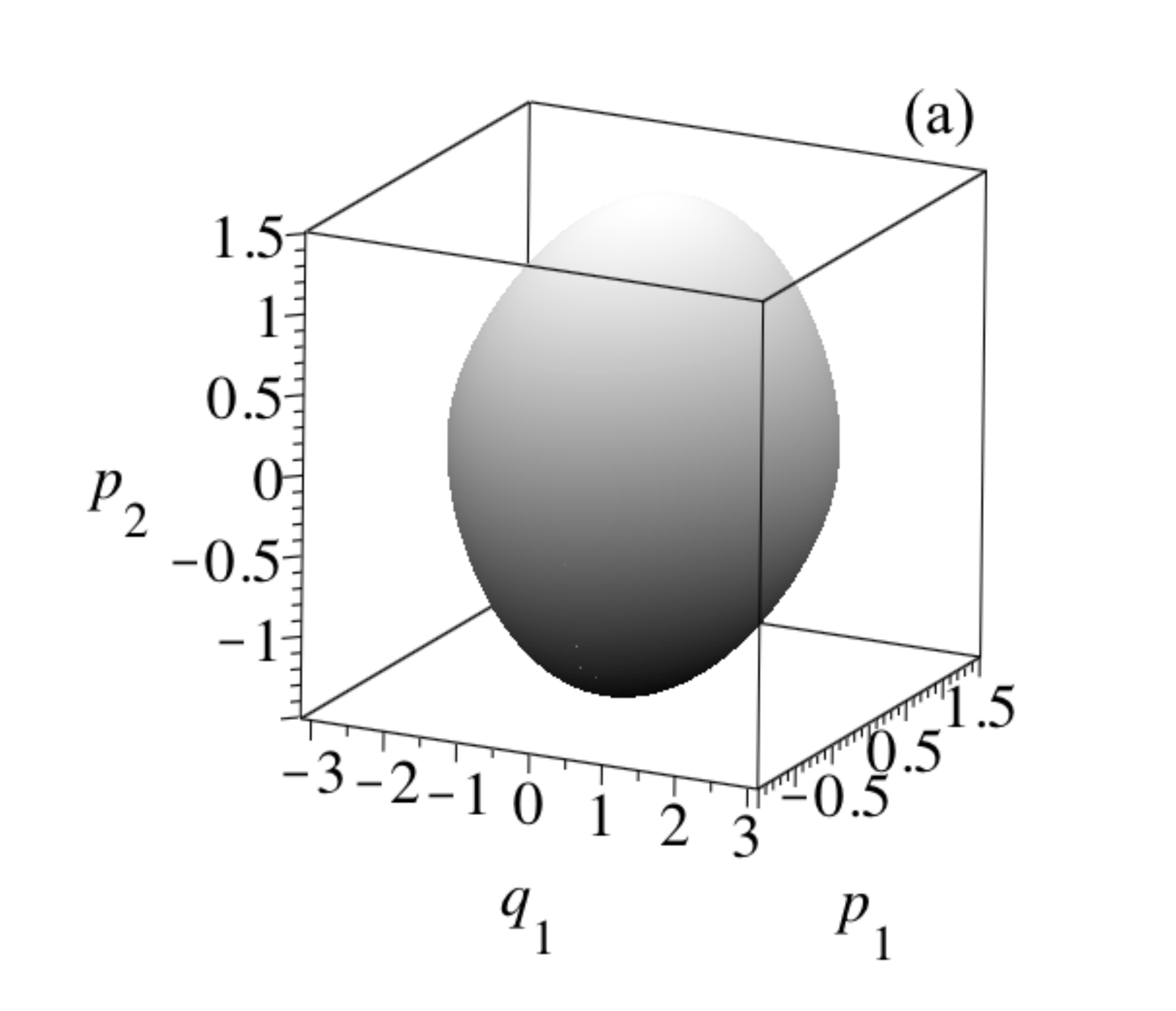} \\
\includegraphics[scale=0.3]{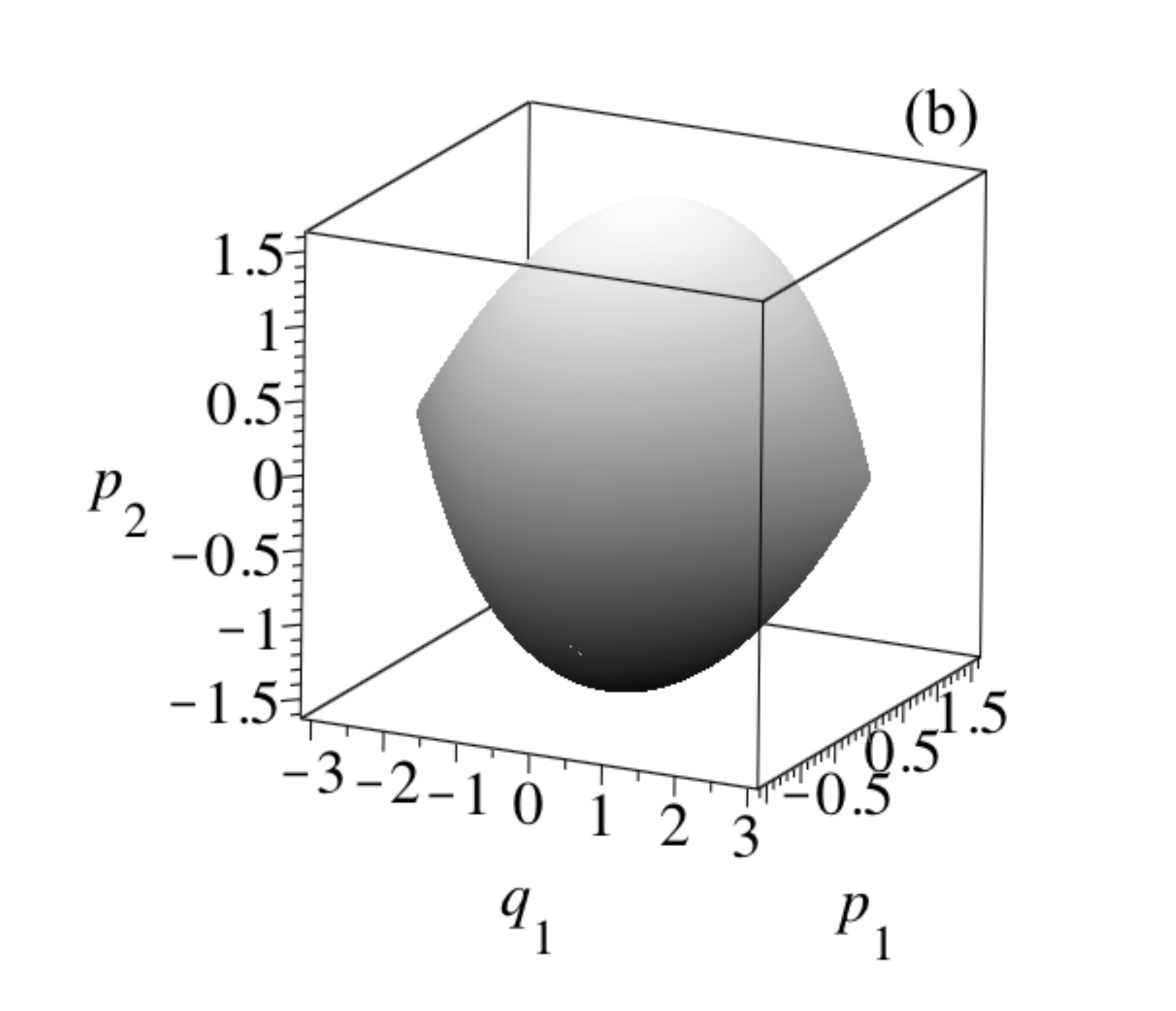} \\
\includegraphics[scale=0.3]{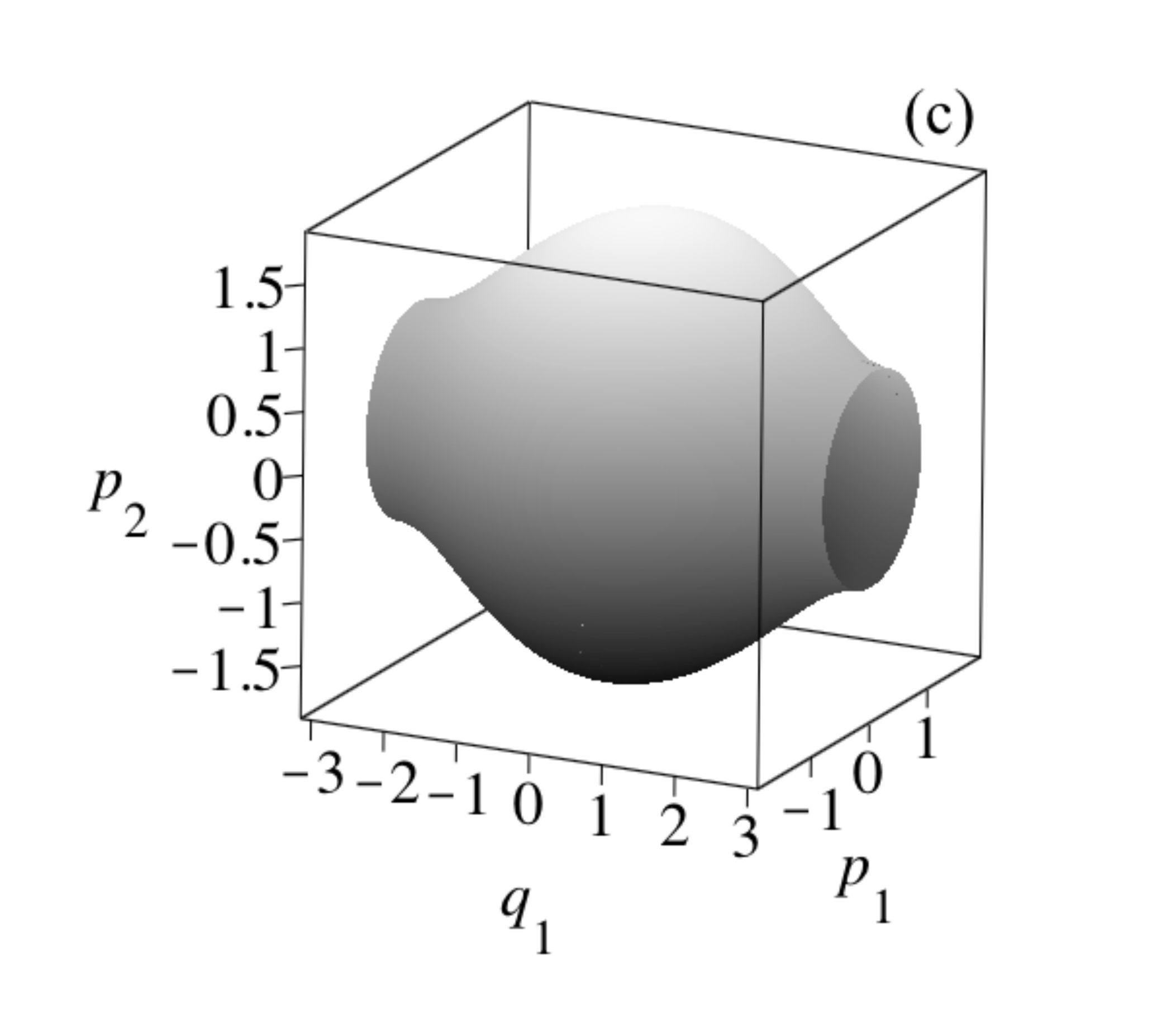}
\end{center}

\vspace*{1.5cm}
FIGURE 3

\newpage

\begin{center} 
\includegraphics[scale=1.0]{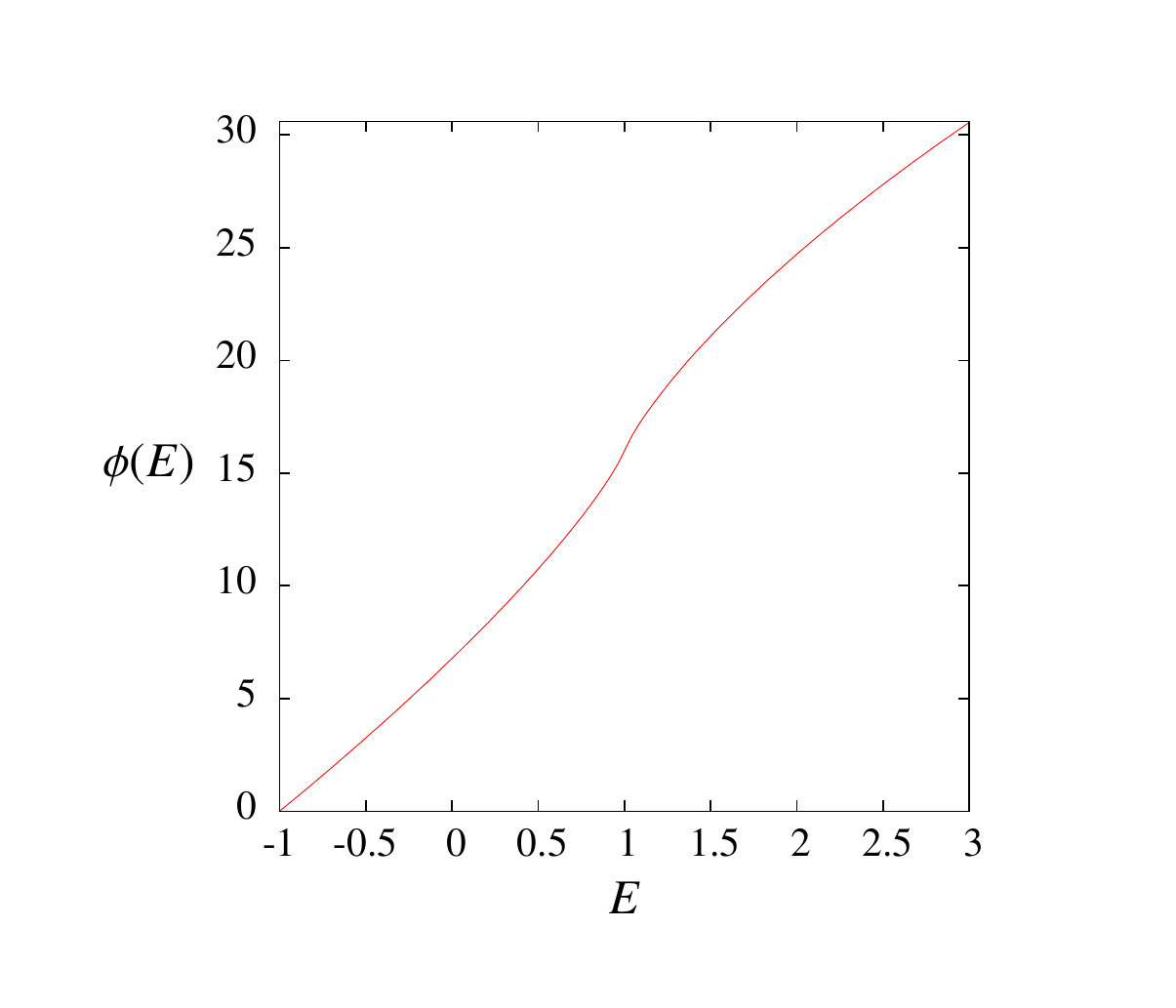}
\end{center}

\vspace*{1.5cm}
FIGURE 4

\newpage

\begin{center} 
\includegraphics[scale=0.3]{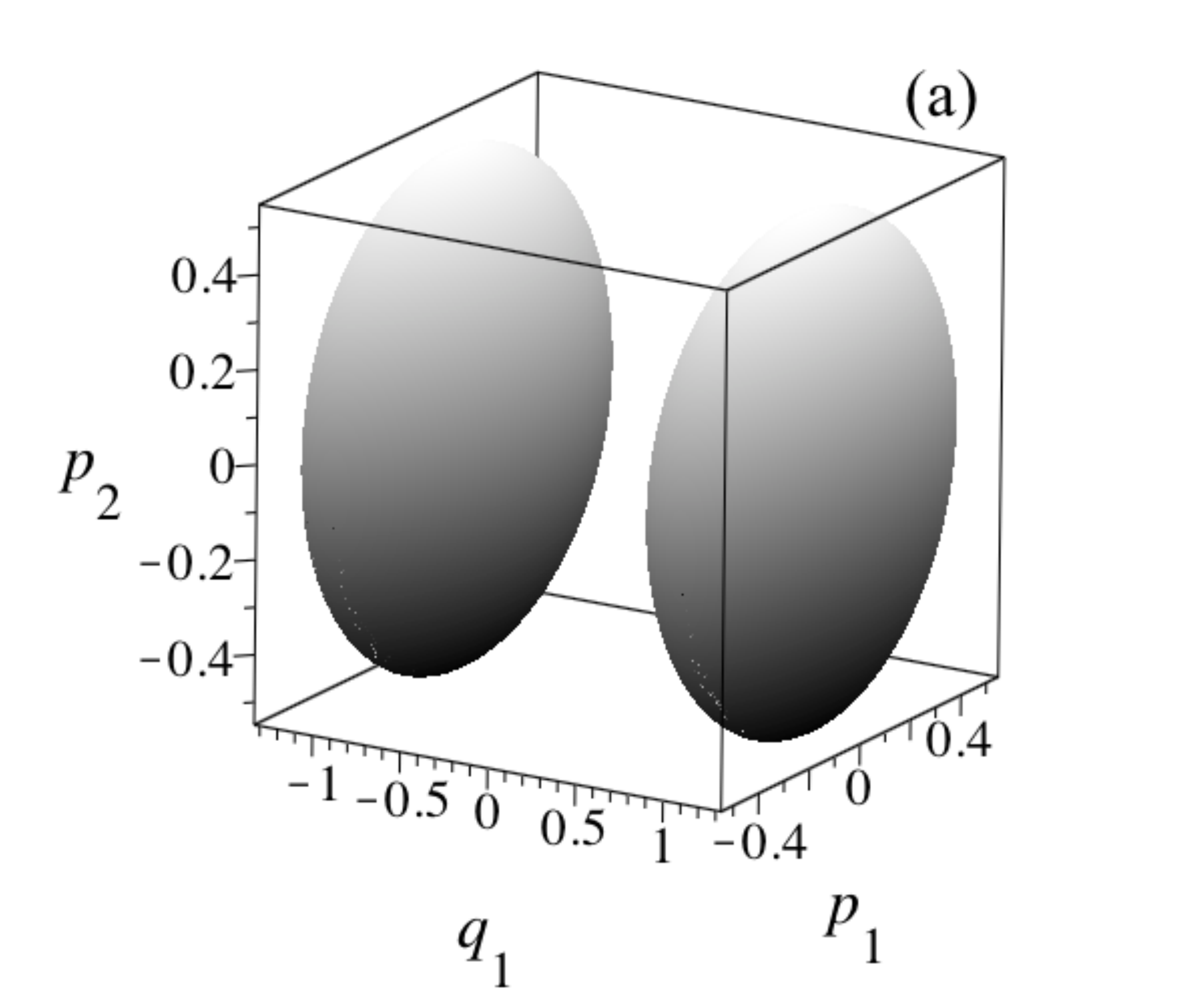} \\
\includegraphics[scale=0.3]{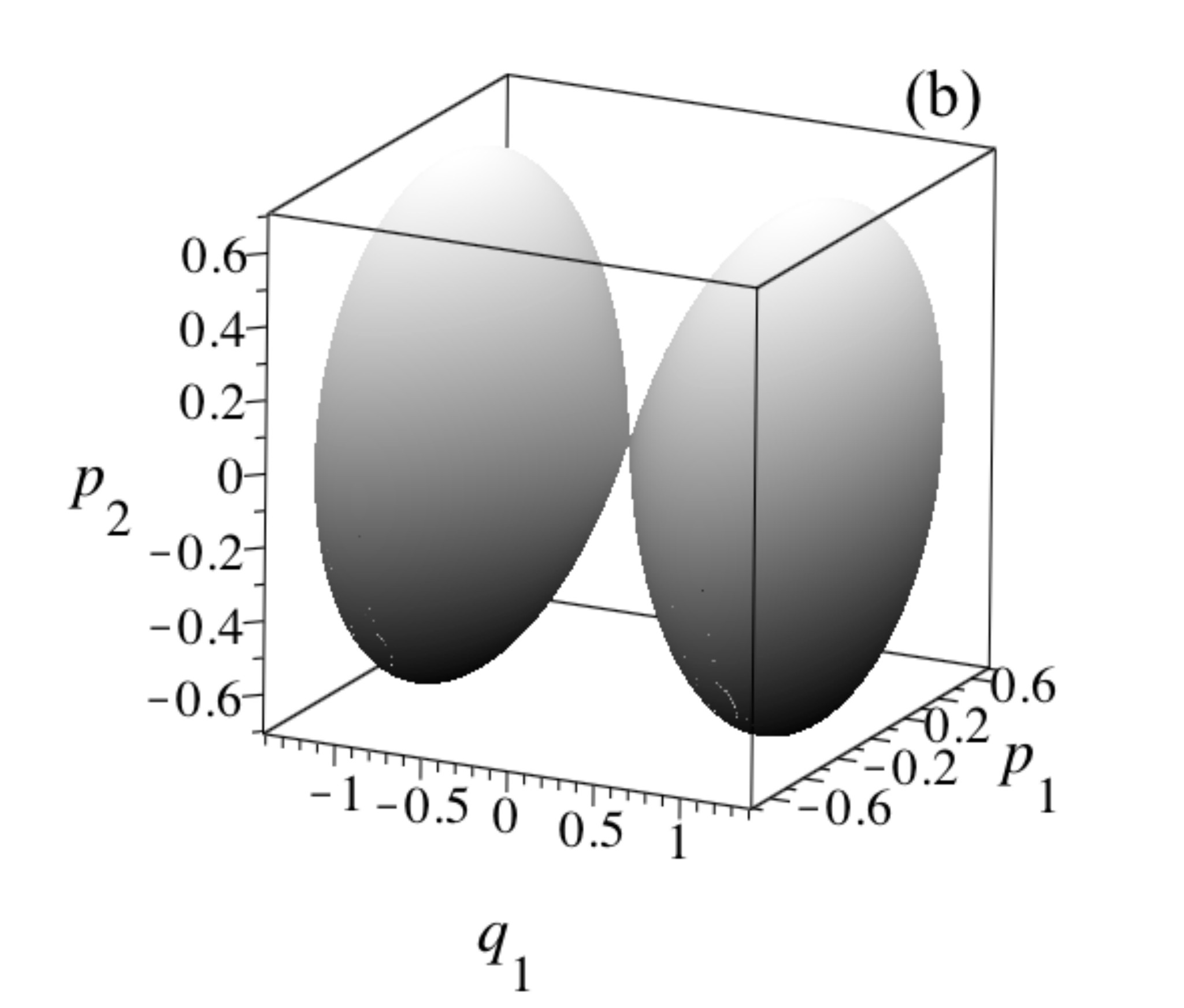} \\
\includegraphics[scale=0.3]{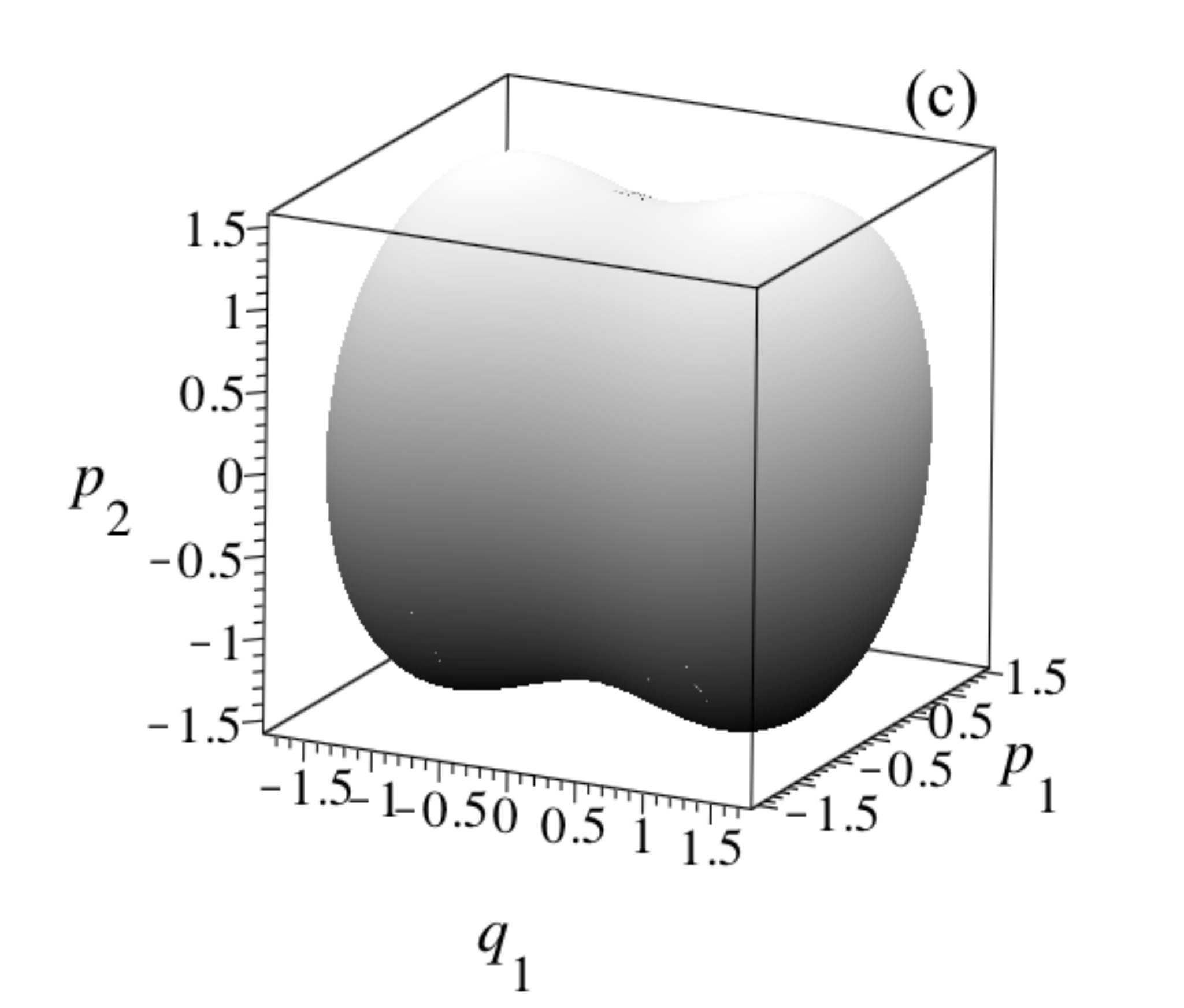}
\end{center}

\vspace*{1.5cm}
FIGURE 5
 
\newpage

\begin{center} 
\includegraphics[scale=1.0]{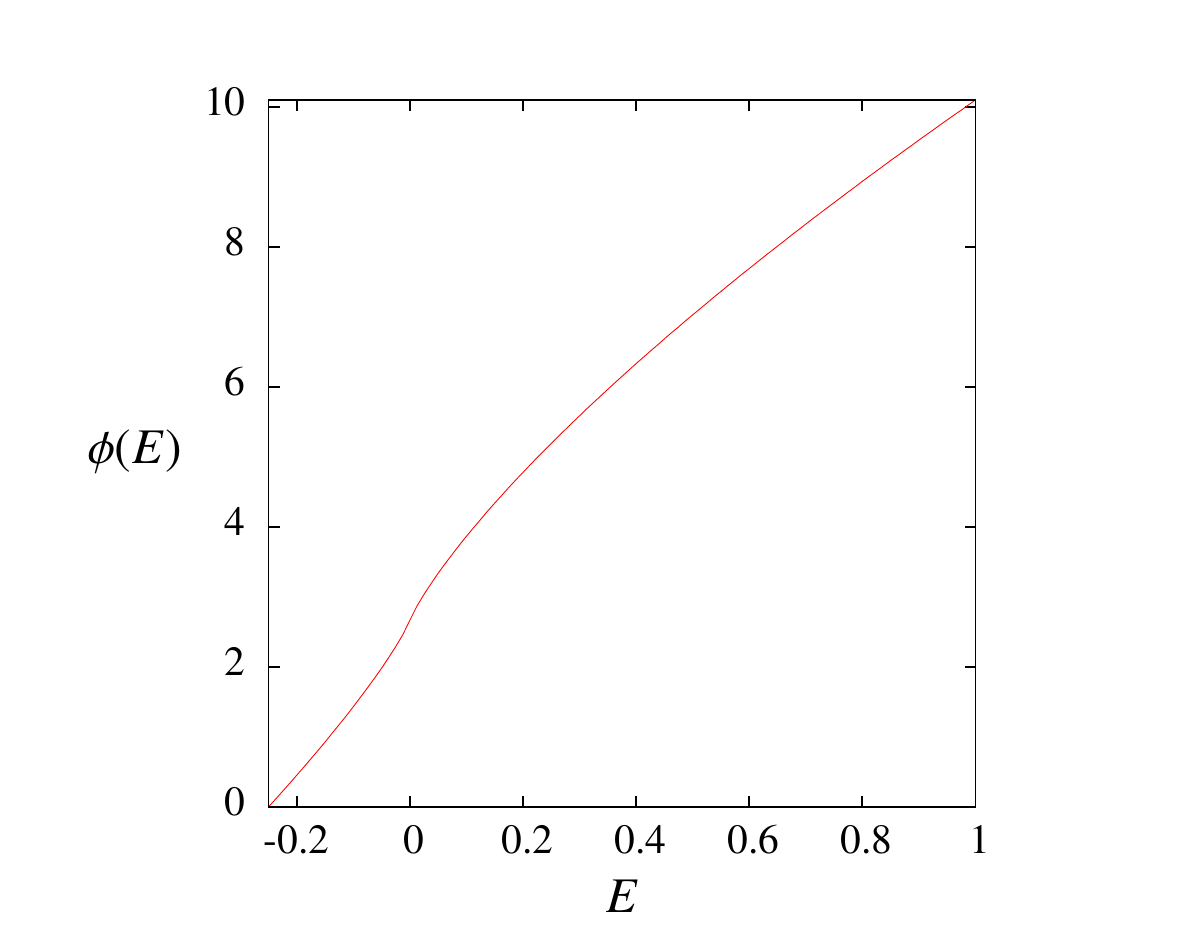}
\end{center}

\vspace*{1.5cm}
FIGURE 6

\newpage

\begin{center} 
\includegraphics[scale=1.0]{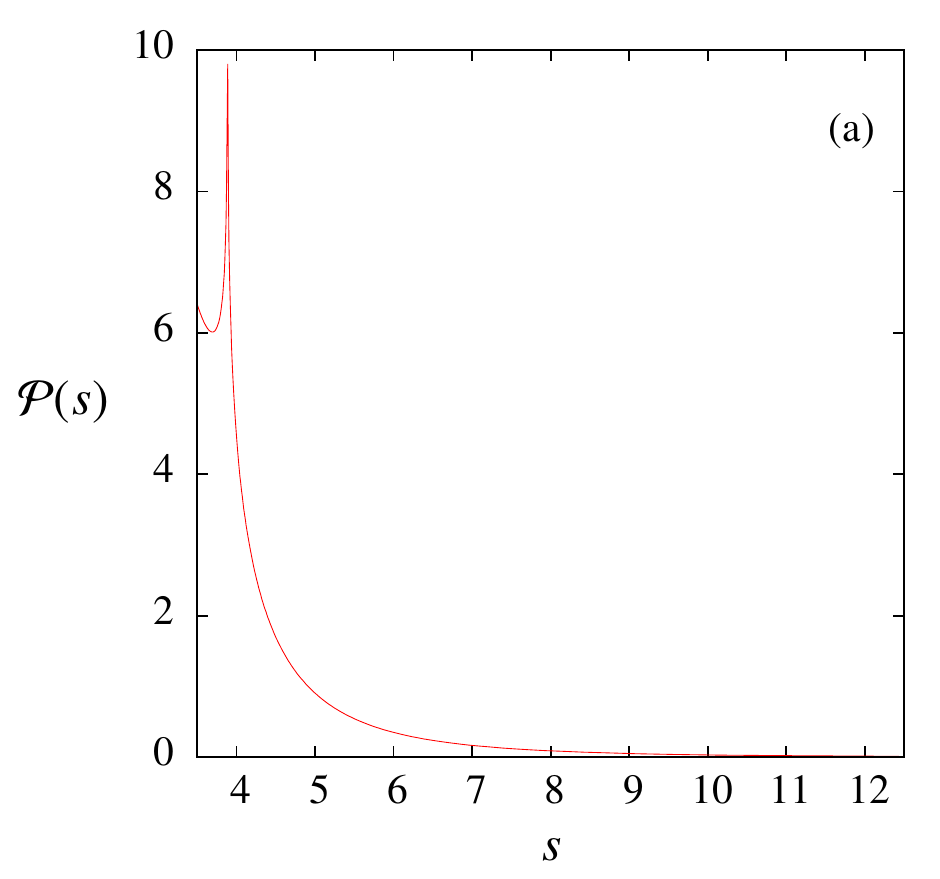}
\includegraphics[scale=1.0]{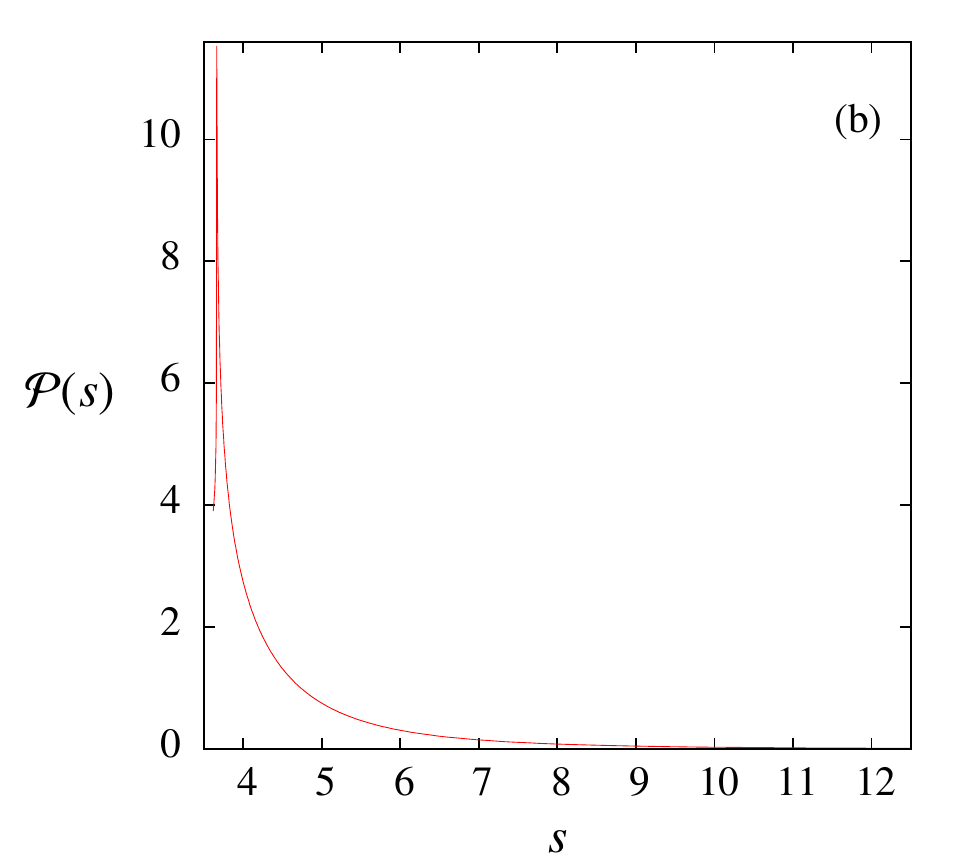}
\end{center}

\vspace*{1.5cm}
FIGURE 7



\begin{thebibliography}{10}

\bibitem{Pechukas81}
P.~Pechukas.
\newblock Transition {S}tate {T}heory.
\newblock {\em Ann. Rev. Phys. Chem.}, 32:159--177, 1981.

\bibitem{PollakTalkner}
E.~Pollak and P.~Talkner.
\newblock {Reaction rate theory: {W}hat it was, where it is today, and where is
  it going?}
\newblock {\em Chaos}, 15:026116, 2005.

\bibitem{LaidlerKing}
K.~J. Laidler and M.~C. King.
\newblock The development of transition state theory.
\newblock {\em J. Phys. Chem.}, 87:2657--2664, 1983.

\bibitem{Garrett}
B.~C. Garrett.
\newblock Perspective on ``{T}he transition state method,'' {W}igner {E}.
  (1938) {Trans.} {F}araday {S}oc. 34:29-41.
\newblock {\em Theor. Chem. Acc.}, 103:200--204, 2000.

\bibitem{Petersson}
G.~A. Petersson.
\newblock Perspective on ``{T}he activated complex in chemical reactions,''
  {E}yring, {H.} (1995) {J}. {C}hem. {P}hys. 3: 107.
\newblock {\em Theor. Chem. Acc.}, 103:190--195, 2000.

\bibitem{Wigner1938}
E.~Wigner.
\newblock The transition state method.
\newblock {\em Trans. Faraday Soc.}, 34:29--41, 1938.

\bibitem{child1980analytical}
M.~S. Child and E.~Pollak.
\newblock Analytical reaction dynamics: Origin and implications of trapped
  periodic trajectories.
\newblock {\em The Journal of Chemical Physics}, 73:4365, 1980.

\bibitem{pollak1980classical}
E.~Pollak and M.~S. Child.
\newblock Classical mechanics of a collinear exchange reaction: A direct
  evaluation of the reaction probability and product distribution.
\newblock {\em The Journal of Chemical Physics}, 73:4373, 1980.

\bibitem{polChilPec1980}
E.~Pollak, M.~S. Child, and P.~Pechukas.
\newblock Classical transition state theory: a lower bound to the reaction
  probability.
\newblock {\em The Journal of Chemical Physics}, 72(3):1669--1678, 1980.

\bibitem{pechukas1979classical}
P.~Pechukas and E.~Pollak.
\newblock Classical transition state theory is exact if the transition state is
  unique.
\newblock {\em The Journal of Chemical Physics}, 71:2062, 1979.

\bibitem{pollak1978transition}
E.~Pollak and P.~Pechukas.
\newblock Transition states, trapped trajectories, and classical bound states
  embedded in the continuum.
\newblock {\em The Journal of Chemical Physics}, 69:1218, 1978.

\bibitem{pechukas1977trapped}
P.~Pechukas and E.~Pollak.
\newblock Trapped trajectories at the boundary of reactivity bands in molecular
  collisions.
\newblock {\em The Journal of Chemical Physics}, 67(12):5976--5977, 1977.

\bibitem{pollak1979unified}
E.~Pollak and P.~Pechukas.
\newblock {Unified statistical model for''complex''and''direct''reaction
  mechanisms: A test on the collinear H + H$_2$ exchange reaction}.
\newblock {\em The Journal of Chemical Physics}, 70(1):325--333, 1979.

\bibitem{Devogelaere55}
R.~Devogelare and M.~Boudart.
\newblock Contribution to the theory of reaction rates.
\newblock {\em J. Chem. Phys.}, 23:1236--1244, 1955.

\bibitem{wiggins90}
S.~Wiggins.
\newblock On the geometry of transport in phase space {I}. {T}ransport in $k$
  degree-of-freedom {H}amiltonian systems, $2 \le k < \infty$.
\newblock {\em Physica D}, 44:471--501, 1990.

\bibitem{Gillilan91}
R.~E. Gillilan and G.~S. Ezra.
\newblock Transport and turnstiles in multidimensional {H}amiltonian mappings
  for unimolecular fragmentation: {A}pplication to van der {W}aals
  predissociation.
\newblock {\em J. Chem. Phys.}, 94:2648--2668, 1991.

\bibitem{Gillilan89}
R.E. Gillilan and W.P Reinhardt.
\newblock {Barrier recrossing in surface diffusion: a phase space perspective}.
\newblock {\em Chemical Physics Letters}, 156:478--482, 1989.

\bibitem{gillilan1990invariant}
R.E. Gillilan.
\newblock Invariant surfaces and phase space flux in three-dimensional surface
  diffusion.
\newblock {\em The Journal of Chemical Physics}, 93:5300, 1990.

\bibitem{wiggins2001impenetrable}
S.~Wiggins, L.~Wiesenfeld, C.~Jaff{\'e}, and T.~Uzer.
\newblock Impenetrable barriers in phase-space.
\newblock {\em Physical Review Letters}, 86(24):5478--5481, 2001.

\bibitem{uzer2002geometry}
T.~Uzer, C.~Jaff{\'e}, J.~Palaci{\'a}n, P.~Yanguas, and S.~Wiggins.
\newblock The geometry of reaction dynamics.
\newblock {\em Nonlinearity}, 15(4):957, 2002.

\bibitem{waalkens2007wigner}
H.~Waalkens, R.~Schubert, and S.~Wiggins.
\newblock Wigner's dynamical transition state theory in phase space: classical
  and quantum.
\newblock {\em Nonlinearity}, 21(1):R1, 2008.

\bibitem{li2006definability}
C.~B. Li, A.~Shoujiguchi, M.~Toda, and T.~Komatsuzaki.
\newblock Definability of no-return transition states in the high-energy regime
  above the reaction threshold.
\newblock {\em Physical review letters}, 97(2):28302, 2006.

\bibitem{Allahem12}
A.~Allahem and T.~Bartsch.
\newblock Chaotic dynamics in multidimensional transition states.
\newblock submitted to Arxiv.org, 2012.

\bibitem{Inarrea11}
M.~Inarrea, J.~F. Palacian, A.~I. Pascual, and J.~P. Salas.
\newblock Bifurcations of dividing surfaces in chemical reactions.
\newblock {\em J. Chem. Phys.}, 135:014110, 2011.

\bibitem{yang09}
D.~G. Yang.
\newblock Breakdown of normal hyperbolicity for a family of invariant manifolds
  with generalized {L}yapunov type numbers uniformly bounded below their
  critical values.
\newblock submitted to Arxiv.org, 2009.

\bibitem{Fenichel1971}
N.~Fenichel.
\newblock Persistence and smoothness of invariant manifolds for flows.
\newblock {\em Indiana Univ. Math. J}, 21(193-226):1972, 1971.

\bibitem{Fenichel1974}
N.~Fenichel.
\newblock Asymptotic stability with rate conditions.
\newblock {\em Indiana Univ. Math. J.}, 23:1109--1137, 1974.

\bibitem{Fenichel1977}
N.~Fenichel.
\newblock Asymptotic stability with rate conditions, ii.
\newblock {\em Indiana Univ. Math. J.}, 26:81--93, 1977.

\bibitem{hirsch1970invariant}
M.~W. Hirsch, C.~Pugh, and M.~Shub.
\newblock Invariant manifolds.
\newblock {\em Bull. Amer. Math. Soc.}, 76(5):1015--1019, 1970.

\bibitem{wiggins1994normally}
S.~Wiggins.
\newblock {\em Normally hyperbolic invariant manifolds in dynamical systems},
  volume 105.
\newblock Springer, 1994.

\bibitem{Delshams2012}
A.~{Delshams}, M.~{Gidea}, and P.~{Roldan}.
\newblock {Transition map and shadowing lemma for normally hyperbolic invariant
  manifolds}.
\newblock {\em ArXiv e-prints}, April 2012.

\bibitem{bolotin2000remarks}
S.~V. Bolotin and D.~V. Treschev.
\newblock Remarks on the definition of hyperbolic tori of hamiltonian systems.
\newblock {\em Regular and Chaotic dynamics}, 5(4):401--412, 2000.

\bibitem{footnote1}
{For our applications this means everywhere, except at possible isolated points
  in the phase space.}

\bibitem{collins2012isomerization}
P.~Collins, G.~S. Ezra, and S.~Wiggins.
\newblock Isomerization dynamics of a buckled nanobeam.
\newblock {\em arXiv preprint arXiv:1206.3929}, 2012.

\bibitem{thiele1962comparison}
E.~Thiele.
\newblock Comparison of the classical theories of unimolecular reactions.
\newblock {\em The Journal of Chemical Physics}, 36(6):1466--1472, 1962.

\bibitem{thiele1963comparison}
E.~Thiele.
\newblock {Comparison of the classical theories of unimolecular reactions. II.
  A model calculation}.
\newblock {\em The Journal of Chemical Physics}, 38(8):1959--1966, 1963.

\bibitem{ezra:164118}
G.~S. Ezra, H.~Waalkens, and S.~Wiggins.
\newblock Microcanonical rates, gap times, and phase space dividing surfaces.
\newblock {\em The Journal of Chemical Physics}, 130(16):164118, 2009.

\bibitem{arnol1989mathematical}
V.~I. Arnol'd.
\newblock {\em Mathematical methods of classical mechanics}, volume~60.
\newblock Springer, 1989.

\bibitem{binney1985structure}
J.~Binney, O.~E. Gerhard, and P.~Hut.
\newblock Structure of surfaces of section.
\newblock {\em Monthly Notices of the Royal Astronomical Society}, 215:59,
  1985.

\bibitem{meyer:3147}
H-D. Meyer.
\newblock {Theory of the Liapunov exponents of Hamiltonian systems and a
  numerical study on the transition from regular to irregular classical
  motion}.
\newblock {\em The Journal of Chemical Physics}, 84(6):3147--3161, 1986.

\bibitem{toller1985theory}
M.~Toller, G.~Jacucci, G.~DeLorenzi, and C.~P. Flynn.
\newblock Theory of classical diffusion jumps in solids.
\newblock {\em Physical Review B}, 32(4):2082, 1985.

\bibitem{mackay1990flux}
R.~S. MacKay.
\newblock Flux over a saddle.
\newblock {\em Physics Letters A}, 145(8):425--427, 1990.

\bibitem{waalkens2004direct}
H.~Waalkens and S.~Wiggins.
\newblock Direct construction of a dividing surface of minimal flux for
  multi-degree-of-freedom systems that cannot be recrossed.
\newblock {\em Journal of Physics A: Mathematical and General}, 37(35):L435,
  2004.

\bibitem{brumer1980time}
P.~Brumer, D.~E. Fitz, and D.~Wardlaw.
\newblock Time delay for bimolecular collisions: Utility of the spectral
  theorem in the classical limit.
\newblock {\em The Journal of Chemical Physics}, 72:386, 1980.

\bibitem{pollak1981classical}
E.~Pollak.
\newblock A classical spectral theorem in bimolecular collisions.
\newblock {\em The Journal of Chemical Physics}, 74:6763, 1981.

\bibitem{waalkens2005efficient}
H.~Waalkens, A.~Burbanks, and S.~Wiggins.
\newblock Efficient procedure to compute the microcanonical volume of initial
  conditions that lead to escape trajectories from a multidimensional potential
  well.
\newblock {\em Physical review letters}, 95(8):84301, 2005.

\bibitem{waalkens2005formula}
H.~Waalkens, A.~Burbanks, and S.~Wiggins.
\newblock A formula to compute the microcanonical volume of reactive initial
  conditions in transition state theory.
\newblock {\em Journal of Physics A: Mathematical and General}, 38(45):L759,
  2005.

\bibitem{slater:1256}
N.~B. Slater.
\newblock New formulation of gaseous unimolecular dissociation rates.
\newblock {\em The Journal of Chemical Physics}, 24(6):1256--1257, 1956.

\bibitem{slater1959theory}
N.~B. Slater.
\newblock {\em Theory of unimolecular reactions}.
\newblock Cornell University Press Ithaca, New York, 1959.

\bibitem{Dumont1986}
R.~S. Dumont and P.~Brumer.
\newblock Dynamical theory of statistical unimolecular decay.
\newblock {\em The Journal of Physical Chemistry}, 90(16):3509--3516, 1986.

\bibitem{carpenter2003nonexponential}
B.~K. Carpenter.
\newblock Nonexponential decay of reactive intermediates: new challenges for
  spectroscopic observation, kinetic modeling and mechanistic interpretation.
\newblock {\em Journal of physical organic chemistry}, 16(11):858--868, 2003.

\bibitem{bunker1962monte}
D.~L. Bunker.
\newblock {Monte Carlo Calculation of Triatomic Dissociation Rates. I. NO and
  O}.
\newblock {\em The Journal of Chemical Physics}, 37:393, 1962.

\bibitem{bunker1964monte}
D.~L. Bunker.
\newblock {Monte Carlo calculations. IV. Further studies of unimolecular
  dissociation}.
\newblock {\em The Journal of Chemical Physics}, 40:1946, 1964.

\bibitem{bunker1973non}
D.~L. Bunker and W.~L. Hase.
\newblock {On non-RRKM unimolecular kinetics: Molecules in general, and CHNC in
  particular}.
\newblock {\em The Journal of Chemical Physics}, 59:4621, 1973.

\bibitem{bunker1966theory}
D.~L. Bunker.
\newblock {\em {Theory of Elementary Gas Reaction Rates}}.
\newblock Pergamon Press, 1966.

\bibitem{bunker1968monte}
D.~L. Bunker and M.~Pattengill.
\newblock {Monte Carlo Calculations. VI. A Re-evaluation of the RRKM Theory of
  Unimolecular Reaction Rates}.
\newblock {\em The Journal of Chemical Physics}, 48(2):772--776, 1968.

\bibitem{Hase76}
W.~L. Hase.
\newblock {Dynamics of Unimolecular Reactions}.
\newblock In W.~H. Miller, editor, {\em {Modern Theoretical Chemistry}},
  volume~2, pages 121--170. Plenum, New York, 1976.

\bibitem{grebenshchikov2003state}
S.~Y. Grebenshchikov, R.~Schinke, and W.~L. Hase.
\newblock State-specific dynamics of unimolecular dissociation.
\newblock {\em Comprehensive Chemical Kinetics}, 39:105--242, 2003.

\bibitem{lourderaj2009theoretical}
U.~Lourderaj and W.L. Hase.
\newblock Theoretical and computational studies of non-rrkm unimolecular
  dynamics.
\newblock {\em J. Phys. Chem. A}, 113(11):2236--2253, 2009.

\bibitem{hancock1917elliptic}
H.~Hancock.
\newblock {\em {Elliptic integrals}}.
\newblock John Wiley \& Sons, inc., 1917.

\end{thebibliography}


\newpage
\end{document}